\documentclass[12pt]{article}
\makeatletter \oddsidemargin  -.1in \evensidemargin -.1in
\newcommand{\singlespacing}{\let\CS=\@currsize\renewcommand{\baselinestretch}{1}\tiny\CS}
\newcommand{\oneandahalfspacing}{\let\CS=\@currsize\renewcommand{\baselinestretch}{1.25}\tiny\CS}
\newcommand{\doublespacing}{\let\CS=\@currsize\renewcommand{\baselinestretch}{1.35}\tiny\CS}

\newtheorem{rule-def}[theorem]{Rule}

\RequirePackage[dvips]{graphicx} \textheight 22.5cm
\usepackage{amssymb}
\usepackage{amsmath}
\usepackage{psfrag}

\begin{document}

\title{\bf Electroosmotic flow of a rheological fluid in non-uniform
  micro-vessels} \author{\small S. Maiti$^{1,2}$ \thanks{Email
    address: {\it Corresponding author,
      maiti0000000somnath@gmail.com/somnath.maiti@lnmiit.ac.in
      (S. Maiti)}}, S. K. Pandey$^{1}$ \thanks{Email address: {\it
      skpandey.apm@itbhu.ac.in (S. K. Pandey)}}, J. C. Misra $^{3}$
  \thanks{Email address: {\it misrajc@gmail.com (J. C. Misra)}}\\ \it
  \it $^{1}$Department of Mathematical Sciences\\ \it Indian Institute
  of Technology (BHU),\\ Varanasi-221005, India\\$^{2}$ Department of
  Mathematics, The LNM Institute of Information Technology\\Jaipur
  302031, India \\$^{3}$ Honorary Visiting Professor, Department of
  Mathematics,\\ Ramakrishna Mission Vidyamandira, Belur Math-711202,
  Howrah, India} \date{} \maketitle \noindent \doublespacing

\begin{abstract}
The paper deals with a theoretical study of electrokinetic flow of a
rheological Herschel-Bulkley fluid through a cylindrical tube of
variable cross-section. The concern of this study is to analyze
combined pressure-driven and electroosmotic flow of Herschel-Bulkley
fluid. The wall potential is considered to vary slowly and
periodically along the axis of the tube. With reference to flow in the
micro-vessels, the problem has been solved using the lubrication
theory. The Helmholtz-Smoluchowski (HS) slip boundary condition has
been employed in this study. Volumetric flow rate $Q$ is found to be
significantly affected by the yield stress parameter $\nu$ only if an
applied pressure force is active. The linear superposition of flow
components separately due to the hydrodynamic and electric force
occurs only for a strictly uniform tube. This linear relationship
fails if non-uniformity appears in either tube radius or in
distribution of the electrokinetic slip boundary condition. Moreover,
converging/diverging nature of the mean tube radius plays a crucial
role on the fluid transport. For the benefit of readers, along with
the original contribution, some applications of external electrical
stimulation (ES) in the human body and HS slip velocity, studied in
the past by previous researchers have been discussed in the
paper.
\end{abstract}

\section{Introduction}
Movement of fluid in a micro-vessel can take place under the action of
electrokinetic forces, or a pressure gradient, or a combination of
both of them. Such a flow is called an electroosmotic flow (EOF). The
flow takes place due to the interaction of an electric field and
mobile ions in the electric double layer (EDL) formed on a charged
surface. EDL consists of an inner Stern layer and an outer diffuse
layer. The former one is formed at the position where counterions are
firmly attached to the surface. The latter one is at the location
where counterions outnumber co-ions and form a charged atmosphere
shielding the bulk solution from the charged surface \cite{Li}. There
are several advantages of EOF over pressure-driven flow. The
advantages include a more precise flow control, a velocity profile
which is much flatter, and flow rate which is less dependent on the
vessel-size.

There has been a growing interest in the studies of electroosmotic
flow in micro-vessels among engineers and scientists in the emerging
areas of micro-fluidics and bio-chips. The mechanism of this flow can
be used in various microfluidic applications including miniaturized
flow injection analysis \cite{Jacobson,Ermakov}, and DNA
amplification, medical diagnostics, chemical analysis, material
synthesis, energy harvesting, clean up, separation, and detection
\cite{Hjerten,Chen}.

As mentioned in scientific literatures \cite{Dutta}, the first
discovery of the electrokinetic effects was made by Reuss
\cite{Reuss,Hunter,Zhao1} by performing experimental investigations on
porous clay and it was followed by experiments of Wiedemann
\cite{Wiedemann}. Relating the electric and flow parameters for
electrokinetic transport, Helmholtz \cite{Helmholtz} produced the
electric double layer (EDL) theory in 1879. M. von Smoluchowski
\cite{Smoluchowski} derived a slip velocity condition for
electroosmotically driven flows when EDL thickness was much smaller
than the channel dimensions. In order to resolve the velocity field
within the EDL, it is essential to place many grid points massively
clustered near the wall which create difficulties in the numerical
solution of electroosmotic flows. The HS velocity is adopted as a
well-known practical alternative to alleviate the above mentioned
difficulties \cite{Probstein}. It is an artificial slip velocity
imposed at the solid surface in order to incorporate the
electroosmotic force in capillaries. For Newtonian fluids, the HS slip
velocity is well documented as well as widely adopted by researchers
working in this area \cite{Park}.

The micro-channels, used on a chip for many laboratory applications,
have a typical height of 1-100 $\mu m$ and the electroosmotic forces
are concentrated within the EDL having an effective thickness in the
order of 1-100 $nm$ \cite{Dutta}. Since for electroosmotically driven
micro-flows, it is a a great challenge in numerical simulation due to
this 2-5 orders of magnitude difference in the EDL and the channel
length scales; it is required to build up a unified slip condition
incorporating the EDL effects by specifying an appropriate
velocity-slip condition on the wall \cite{Dutta}. It is generally
accepted that task of resolving the double layers is directly
computationally impractical \cite{Patankar,Mitchell} for channel
widths of the order of tens of microns, since the EDL thickness is
then at least two orders of magnitude smaller than the size of the
computational domain itself \cite{MacInnes}. In fact the order of
magnitude is less than about 100 $nm$.

The HS slip velocity boundary condition is generally used in
computational models of flows through micro-channels, because it offers
the advantage to approximate the motion of the EDL without resolving
the charge density profiles close to the walls \cite{Craven}. Owing to
this, the time of computation is drastically reduced and so the flow
equation can be easily solved. Of course, in the region where the wall
curvature is sufficiently high, the magnitude of nonphysical velocity
is quite large. Keeping in mind that the hydraulic radius varies
between 10 $\mu m$ and 100 $\mu m$, the thickness of the EDL is must
less than that of channel, in the case of microfluidic devices. Thus
the non-dimensional electrokinetic radius is considerably
large. Because of this, the problem of determining the electrical
double-layer potential through the use of numerical method turns out
to be highly ill-conditioned, whereby determination of different
physical quantities becomes very difficult \cite{Xuan}. In order to
avoid the calculation of electrical double-layer, an electroosmotic
slip condition is generally taken as an alternative to simplify the
direct current (DC) electroosmotic flow field
\cite{Ghosal,Sinton,Xuan2,Bianchi,Erickson}. Some authors have used
the slip condition in different studies of AC electroosmosis
\cite{Ajdari,Oddy,Dutta2}.

As hypothesized by Ermakov et al. \cite{Ermakov,Ermakov2}, one can
represent the effect of the double layer by substituting the no-slip
condition at the channel walls with the velocity outside the double
layer in the classical one-dimensional electroosmotic flow solution
(e.g. Hunter \cite{Hunter1,Hunter2}) in order to reduce the
computational time substantially. As examined by Dutta and Beskok
\cite{Dutta}, the bulk velocity field extended up to the wall had a
constant slip value equivalent to Helmholtz-Smoluchowski (HS)
slip-velocity. The authors reported that the appropriate slip velocity
on the wall was the HS velocity even for finite EDL thickness
conditions. It is usually considered in the case of simple
micro-channel flows, the EDL at walls is thin enough for ``slip
velocity'' boundary conditions to be employed with sufficient accuracy
\cite{MacInnes}. MacInnes \cite{MacInnes2} used the limits of this
approach theoretically in cases characterized by non-uniform liquid
properties and complex channel geometries considering the chemically
reacting flow through any channel of arbitrary geometrical
configuration. The flow is supposed to be produced by electric
potential and pressure differences with heat transfer and
electrophoresis \cite{MacInnes}.

For channel widths in the range from several microns to hundreds of
microns, MacInnes \cite{MacInnes2} reported that in the case of local
one-dimensional solutions to the Poisson-Boltzmann equation, using
appropriate boundary conditions on the walls the expressions for the
double layer agree with those obtained by previous
researchers. Instead of resolving the EDL directly, it can be replaced
by a slip boundary condition for velocity which gives equivalent
boundary conditions for the core flow
\cite{Ermakov2,MacInnes2,Zimmerman}. With consideration of a Debye
length of less than 10 $nm$ and channel widths of the order of 100
$\mu m$, they \cite{Zimmerman} reported that the layer model
approximation was excellent for the flow regimes considered in
\cite{MacInnes2}.

Some conditions for the validation of the approximation
\cite{MacInnes2} are that (a) the magnitude of the applied electric
field $E=|\mathbf{E}|$ is small compared to the product of the
electrostatic field in the EDL ($\psi_0 /\delta$) and the ratio of the
channel to EDL length scales ($L/\delta$)
i.e. $E<<(\psi_0/\delta)(L/\delta)$ although the condition is spoiled
when the non-dimensional electric field strength rises beyond about
ten times its value in a straight channel \cite{MacInnes2,Craven} and
(b) $\kappa^2 L_0^2>>$max$[1, (E_0L_0 /\zeta_0)^2 ]$ if $L_0$ be the
channel width, $E_0$ the electric field strength, $\zeta_0$ the wall
zeta potential and $\kappa$ the inverse of the Debye length, which
represents the characteristic thickness of the double layer
\cite{MacInnes2,Zimmerman}.

In cases where the double layer thickness is much smaller than the
channel size, it is well known \cite{MacInnes,Probstein} that the
developed flow in a channel may be reckoned using the HS ``slip
velocity'' boundary condition. This equivalent boundary condition
approach is used in the determination of various steady and switching
flows in two-dimensional micro-channel junctions
\cite{Ermakov,Ermakov2,MacInnes}. The electroosmotic potential, in the
limit of small yet finite Debye layers, loses strength very fast
within the thin EDL and a uniform ``plug like'' velocity profile can
be observed in most of the channel \cite{Dutta}. The plug flow
behavior has been found in various experiments
\cite{Molho,Paul,Herr}. The fluid in a capillary travels like plug
flow under the action of electroosmotic force as the velocity at the
solid surface can be considered to be nonslip and the driving force
for fluid movement is acting only within the thin electric double
layer. At the limit of very thin double layer, the velocity slips at
the wall and it changes from a uniform finite velocity to zero
discontinuously at the wall. The velocity deteriorates continuously
across the layer and becomes zero at the wall for an actual
finite-thickness EDL \cite{Park}.

Projecting a non-dimensional time scale $\bar{\omega}$, the ratio of
the diffusion time of momentum across the electric double-layer
thickness to the period of the applied electric field in typical
microfluidic applications, Xuan and Li \cite{Xuan} analytically
demonstrated that the HS electroosmotic velocity is an appropriate
slip condition for alternating current (AC) electroosmotic flows. In
the range of applied frequency $0~Hz<f<16~MHz$, the slip condition
approach is appropriate for AC electroosmotic flows in typical
microfluidic applications when $\bar{\omega}<0.01$ \cite{Xuan}. Due to
the thin EDL in most microfluidic applications, we can get a small
$\bar{\omega}$ ensuring that the HS electroosmotic velocity is an
appropriate slip condition \cite{Xuan}.

However, neglecting the electrical body force in the Cauchy momentum
equation and reducing an EOF flow problem to just a boundary condition
at surface can reduce the validation of results to limited cases with
narrow ranges for physical parameters. Just as an example, in
microfluidic problems with small bulk ionic number concentration, this
assumption may be not valid anymore. On the other hand, in the pure
pressure-driven electrokinetics flow problems (absence of electric
force), other physical phenomena such as streaming potential and
consequently conduction current are dominant where the employing of
the assumption cannot be reasonable.

As considered by Rubin et al. \cite{Rubin}, the tangential components
on both the boundaries are subject to either Helmholtz-Smoluchowski
(HS) slip or Navier slip boundary conditions. HS slip appears over
electrically charged surfaces in the presence of an externally applied
tangential electric field. Because of the interaction of this field
with the excess of net charge in the electric double layer (EDL), the
movement of fluid outside the outer edge of the EDL is taking place
according to the HS equation \cite{Hunter1,Hunter2}. Rubin et al.
\cite{Rubin} have assumed a low Dukhin number for their study such
that the ionic species concentrations in the bulk and the associated
electric field are uniform outside the EDL. Otherwise, non-homogeneity
can be introduced into the resultant electric field in the bulk by
non-homogeneities of zeta potential and surface conduction
\cite{Yariv,Khair}. Hydrophobic surfaces may also create slip which is
typically modelled by a Navier boundary condition describing that the
slip velocity near a flat surface is proportional to the local
velocity gradient \cite{Rubin}. Experiments and molecular dynamics
simulations over the past few decades have indeed corroborated that
slip appears in pressure-driven flows over smooth solvophobic surfaces
with slip lengths of the order of nanometres
\cite{Vinogradova,Baudry}.

One may be curious to know how it is possible to apply electric force
in the micro-vessel and people normally think that electricity flowing
through bodies as an unusual occurrence. As example, they may consider
rather unique animals, such as electric eels, or rare events such as
being struck by lightning. But, most people cannot realize that
electricity is part of everything their body does as from thinking to
doing aerobics and even sleeping. Some applications of electric
stimulations that can be applied to a human are discussed below.

Transcranial direct electric stimulation (tDCS) is applied in humans
by using direct current over the scalp with the help of electrodes. It
was reported that stimulation of cerebellar neurons increased diameter
of both adjacent arterioles and the upstream vessels
\cite{Iadecola,Pulgar}. The reports are also demonstrating the
propagation of vascular responses induced by enhanced neural activity
\cite{Iadecola,Pulgar}. As reviewed by Pulgar \cite{Pulgar}, tDCS may
also modulate blood flow in subcortical structures
\cite{Lang,Nonnekes} together with the demonstration of broader
effects of tDCS on cerebral blood flow (CBF). As reviewed by Thakral
et al. \cite{Thakral}, electrical stimulation (ES) therapy enhances
venous flow \cite{Velmahos,Doran,Doran2,Doran3} in addition to
increased skin perfusion. Transcutaneous electrical nerve stimulation
(TENS), consists of a battery-powered device that delivers electrical
impulses through electrodes, is performed to evaluate the effect of ES
on perfusion.

As mentioned by Rajendran et al. \cite{Rajendran}, in order to enhance
wound healing, the ES is commonly applied by placing electrodes around
the wound, which then deliver short bursts of electrical potential
having result in electrical currents \cite{Ud-Din,Hampton,Arora,Kloth}
[in Table 2 \cite{Ud-Din,Kloth}]. Watanabe et al. \cite{Watanabe}
reported that for electrical recording and stimulating, a minimally
invasive technique having ability of simultaneously monitoring the
activity of a significant number (e.g., $10^3$ to $10^4$) of neurons
is an absolute prerequisite in developing an effective brain-machine
interface. Action potentials can be elicited using a constant current
pulse delivered through a bipolar stimulation electrode placed on the
spinal cord.

The result, subcontractile ES induces enhanced vascularization in
animals, has suggested for one possible route for noninvasive
induction of micro-vessels for augmentation of micro-vessel number in
patients with peripheral vascular disease \cite{Clover}. An Elpha 4
Conti transcutaneous electrical stimulator (ERP Group Ltd, Laval,
Quebec, Canada) has been applied to deliver localized stimulation at
10 $mA$, 1.0 $V$ at a frequency of 8 $Hz$. Electrodes were used to a
5$\times$5-$cm$ area on the first metatarsal joint of the affected
limb and for the 6-week treatment period, stimulation was delivered
for three equally spaced one hour periods each day \cite{Clover}.

Action potentials may be generated through cortical neurons by direct
electrical stimulation (DES). DES may induce synaptic release of
various neurotransmitters that can further react with nearby
astrocytes, smooth muscle cells and endothelial cells of the vessel
wall \cite{Allen,Tsytsarev}. As a result of these reactions, cortical
vessels are driven to either vasoconstriction or vasodilatation from
the resting state but on the other hand, smooth muscle cells and
astrocytes can directly respond to ES \cite{Tsytsarev2}: DES of smooth
muscle cells may induce vasoconstriction, while DES of astrocytes may
drive intracellular calcium waves along the astrocytic syncytium,
which may signal the release of various neuromodulators to direct
blood vessels in order to regulate the metabolic supply by changing
the local vessel capacity. In addition, ES of brain tissue temporarily
enhances extracellular potassium concentration potentially stimulating
neural hyperactivity. Enhanced neural activity needs more blood flow,
which may activate vasodilatation \cite{Allen,Zonta}. Both
vasodilatation and vasoconstriction in response to cortical ES can be
observed using OR-PAM \cite{Tsytsarev}. For the induction of the
preset ES [\cite{Tsytsarev}, Figs. 1(b) and 1(c)], a monopolar
tungsten electrode (impedance: 1 M$\Omega$; tip diameter: 10 $\mu m$;
MicroProbes for Life Science, Gaithersburg, Maryland) was applied into
the cortex to a depth of 0.1 to 0.2 $mm$ through the opening at about
2 $mm$ lateral to sagittal suture and 2 $mm$ posterior to bregma.

As available in \cite{Choi}, clinical evidence indicates that by
enhancing muscle power and function \cite{Jones,Hong} as well as
raising peripheral blood flow \cite{Willand}, neuromuscular electrical
stimulation (NMES) through transcutaneous or implantable (i.e.,
portable) electrodes has beneficial effects on recovery from
denervation injuries. As observed by Loaiza et al. \cite{Loaiza},
increases in muscle blood flow (MBF) and mean arterial pressure
(47$\pm$ 10$\%$ and 18 $\pm$ 5$\%$) over the baseline can be produced
by applying psilateral ES (5 V, 20 Hz, for 30 s), however there was no
significant changes in MBF in the contralateral muscle.

The arteriolar diameter was observed to rise by 38.9 $\pm$ 5$\%$
following ipsilateral ES \cite{Loaiza}. A hook electrode can be
applied directly around the left saphenous nerve and ES (0.5 ms, 20
Hz, 1, 3 and 5 V for 30 s) can be delivered using an electrical
stimulator (SEN-7203, Nihon Kohden) \cite{Loaiza}. To improve blood
circulation in diabetic peripheral vessels and potentially treat
diabetic peripheral neuropathy, a focused ultrasound (FUS) technique
was developed by Tan el al. \cite{Tan} for effective treatment methods
for diabetic peripheral neuropathy. The transducer was driven by in
all experiments, a function generator (33 521A, Agilent, Santa Clara,
California, USA) and a power amplifier (1040 L, Electronics and
Innovation, Rochester, New York, USA) drove and two ring silver wires
(UL1423 28 AWG B28-1000, AA electronic Test, USA) were used on the
middle three digits as stimulating electrodes to apply an ES with a
pulse width of 0.1 ms and a supramaximal intensity \cite{Tan}.

Since the Study of Reuss \cite{Reuss}, there are numerous
investigations on EOF in the literature. However, studies on EOF of
non-Newtonian fluids were limited until recent years. Most
physiological fluids behave as a non-Newtonian (rheological)
fluid. Many bio-fluids such as blood, protein solutions, DNA
solutions; polymeric solutions, and colloidal suspensions are complex
fluids, which cannot be treated as Newtonian fluids
\cite{Berli1,Berli2,Bandopadhyay,Zimmerman,Misra1,Misra2,Maiti1,Maiti2}. These
fluids are usually investigated by scientists, engineers and
industrialists in microfluidic devices. Therefore, the need for the
study of electroosmotic flow of non-Newtonian fluids has been
increasing during recent years. These fluids can be classified
according to their rheological natures as shear thinning/thickening,
viscoelastic, viscoplastic, structuralized fluids etc.

For Newtonian fluids, the relationship between the driving forces and
the flow rate is normally linear. This linear relationship and the
so-called Onsager relations are not expected to followed by
non-Newtonian fluids due to the nonlinear fluid rheology. However,
considering the effects of wall depletion, Berli and Olivares
\cite{Berli1} have reported a linear summation of the parts separately
due to the hydrodynamic and electric forcings of EOF of a
non-Newtonian fluid. It has been further reported that nonlinear
effects are limited to the pressure-driven component of the flow and
the Onsager reciprocity, which is described as equality of the
non-conjugate streaming coefficients, has been satisfied. However,
they studied one-dimensional flow in uniform micro-channels only and
thus opened up for an unanswered question if such a linear
relationship holds for multi-directional flow.

As described in the articles by Barnes \cite{Barnes}, Tuinier and
Taniguchi \cite{Tuinier} etc., the depletion or skimming layer is a
thin layer near a solid wall. In this layer, the fluid is depleted of
the macro-molecules which constitute the non-Newtonian behavior of the
bulk fluid. Thus, the nature of the fluid in this layer is basically
Newtonian solvent. Since it is not accessible to the center of the
macro-molecules, its thickness can be comparable with the radius of
gyration of the macro-molecules. Berli and Olivares \cite{Berli1}
calculated the thickness of this layer. It was enough thick that can
cover the EDL and hence the layer may confine the electrokinetic
driving force to a portion having the fluid of Newtonian nature. The
hydrodynamic driving force is the only force functioning on the bulk
non-Newtonian fluid. Berli and Olivares pointed out \cite{Berli1} that
the streaming conductance depends only on the Newtonian solvent
viscosity and it is independent on the non-Newtonian rheology of the
bulk fluid. So, if the fluid is fully Newtonian, Onsager reciprocity
can be obtained. Even though the inclusion of depletion layer in the
studies on EOF of non-Newtonian fluids had been done by many
researchers (cf. Berli \cite{Berli2}, Bandopadhyay and Chakraborty
\cite{Bandopadhyay}, Zimmerman et al. \cite{Zimmerman} and Olivares et
al.  \cite{Olivares}), the possible nonlinear interplay between the
hydrodynamic and electric forcings for a non-parallel flow has not
been explored in details. Keeping sight into the flow of bio-fluids in
micro-systems, Zimmerman et al. \cite{Zimmerman}, Das and
Chakraborty \cite{Das} formulated theories on non-Newtonian
electrokinetic flow and transport of non-Newtonian fluids in
micro-channels.

The power-law model explains shear thinning and shear thickening
behaviors of a fluid. But in order to describe the viscoplastic
behaviors of a fluid, a yield stress should be introduced. Among the
various types of non-Newtonian models, Herschel-Bulkley fluid model
\cite{Herschel} is a more general choice for EOF of fluid which
require a finite stress, known as yield stress, in order to
deform. Because, this model has advantage over others since the
corresponding results for fluids represented by Bingham plastic model,
power law model and Newtonian fluid model can be derived as different
particular cases from those of the Herschel-Bulkley fluid
model. Moreover, Herschel-Bulkley fluid model is expected to give more
accurate results than many other non-Newtonian models. Although
Newtonian and several non-Newtonian models have been taken into
consideration for the study of blood circulation, it is realized
(Scott-Blair and Spanner \cite{Scott-Blair}) that Herschel-Bulkley
model describes the behaviour of blood very closely. The magnitudes of
yield stress in different fluids are different. If the stress is
smaller in magnitude than the yield stress, material with a yield
stress either does not move or is transported like a rigid body. The
material flows with a non-linear stress-strain relationship either as
a shear-thickening fluid or a shear-thinning fluid when the yield
stress is exceeded. Some examples of fluids behaving in this fashion
include food products, paints, plastics, slurries, pharmaceutical
products gels, drilling fluids etc. \cite{Nallapu,Moreno}. As reported
in \cite{Das2}, Herschel-Bulkley fluids can be used to represent a
wide range of fluids such as greases, starch pastes, colloidal
suspensions, blood etc. The Herschel-Bulkley fluid model can be also
used for simulating debris flow ( \cite{Moreno,Minatti,Remaitre}).

As the velocity profiles in the arterioles having a diameter less than
0.1 $mm$ examined and explained by Iida \cite{Iida}, blood can be
treated fairly well by a Casson fluid model as well as by the
Herschel-Bulkley model. However, Scott-Blair and Spanner
\cite{Scott-Blair}) observed that blood obeys the Casson equation only
in a limited range, but not at very high and very low shear
rates. Moreover, there is no difference between the Casson fluid model
and Herschel-Bulkley fluid model of the experimental data over the
range where the Casson model is valid. It has been reported that the
Casson fluid model can be applied for moderate shear rates in smaller
diameter tubes although the Herschel-Bulkley fluid model can be
considered at still lower shear rate flow in very narrow arteries
where the yield stress is high \cite{Vajravelu}. The Herschel-Bulkley
fluid model contains one more parameter than the Casson fluid model
does which is expected to provide more detailed information about the
blood and other material properties by the use of the Herschel-Bulkley
model \cite{Vajravelu}.

There are several natural and industrial yield stress
fluids. These includes normal blood with 40$\%$ hematocrit (which has
yield stress 4 $mPa$) \cite{Merrill} and a mineral suspension with 52
wt. $\%$ solids (having yield stress 7 to 11 $Pa$) \cite{Nguyen}. Bird
et al. \cite{Bird} carried out a detailed review of viscoplastic
materials with a yield stress. These models are explaining the
movements of many yield-stress materials such as pastes, slurries, and
suspensions which frequently encountered in industrial
problems.

After a lack of effective review for $\sim$ 30 years, since the
detailed review by Bird et al. \cite{Bird,Frigaard}, there have
recently been a number of studies that provide the breadth of the
field \cite{Balmforth,Coussot,Bonn} dealing with useful introductions
\cite{Huilgol,Ovarlez} and the wonderful 100-year Bingham
commemorative collection of researches
\cite{Cloitre,Coussot2,Ewoldt,Frigaard2,Malkin,Mitsoulis,Saramito}
including targeted reviews and finally the nicely written article by
de Souza Mendes and Thompson \cite{de}. A concise summary such as the
review by Bird et al. \cite{Bird} also no longer possible since there
has been an enormous expansion of the literature.

The basic and fundamental feature of a viscoplastic fluid is the
presence of plugs, namely, at the regions where the stress is below
the critical threshold and where the fluid travels like a rigid body.
Plugs may happen in the interior of the motion or may adhere to the
boundary \cite{Fusi}. Viscoplastic fluids are observed relatively
frequently in applications within a thin-film/lubrication/Hele-Shaw
setting \cite{Fusi}. In lubrication flows, the singularity of the
constitutive equation of a visco-plastic fluid may direct to an
apparent inconsistency referred to as the lubrication paradox
\cite{Putz} indicating to the existence or non-existence of true
unyielded plug regions within the flow \cite{Lipscomb}. Indeed, the
scaling techniques may forecast a plug velocity that varies in the
flow direction so that the rigid region is not truly unyielded in
classical lubrication \cite{Fusi}. Many strategies have been proposed
in recent years to overcome the paradox
\cite{Frigaard3,Muravleva,Putz,Wilson,Fusi4} based on assuming the
existence of a pseudo-plug placed between the unyielded and yielded
regions in which the corrections up to the first order are taken into
account \cite{Muravleva2}. The transition of yielded/unyielded happens
in a tiny neighborhood of the pseudo-yield surface \cite{Fusi}. The
paradox can also be overcomed by assuming that the unyielded part may
undergo small deformations \cite{Fusi2,Fusi3,Fusi5}.

The asymptotic analysis of thin viscoplastic films, or viscoplastic
lubrication theory, are belong to an earlier time during the 1950s and
even comes out in classical texts in 1964 by Pinkus and Sternlicht
\cite{Pinkus}. But, objections to this theory appeared in the 1980s
with the observation that lubrication analyses were apparently
inconsistent \cite{Balmforth}. As reported by Lipscomb and Denn in
1984, the ``lubrication paradox'' highlighted how the theory predicted
plug regions that were apparently below the yield stress although the
velocity field there was not fully rigid \cite{Lipscomb}. This
inconsistency motivated many researchers to study regularized
viscoplastic constitutive models and explore unphysical distinguished
limits. However, such diversions ultimately turn out unnecessary since
the paradox can be resolved by recognizing that the lubrication theory
is the leading-order term of an asymptotic expansion
\cite{Balmforth}. Moreover, an exploration of higher-order terms
provides evidence that the problematic plugs are actually slightly
above the yield stress, allowing for their deformation
\cite{Putz,Walton}. The regions have been designated with a term
pseudoplugs to emphasize their nature and with their borders referred
to as fake yield surfaces. Hence, as a result , there is no
lubrication paradox \cite{Balmforth}.

Limited number of studies of EOF of fluids with a yield stress have
been carried out by researchers \cite{Berli1,Liu,Tang,Ng}. We have
motivated to know how the yield stress will affect an EOF of a
rheological fluid in micro-vessels (which remains largely unknown so
far). Since the effect of yield-stress varies qualitatively and
quantitatively depending on the viscoplastic models, the concern of
this study is to explore the decreasing effect due to a yield stress
on the velocity of EOF of a Herschel-Bulkley fluid. For this model,
analytical expressions have been derived for the velocity which
consists of two regions namely sheared region and non-sheared
region. Sheared region is the area where the shear stress exceeds
yield stress, while for the non-sheared the opposite is true. The
position of the interface, known as the yield surface, separating the
sheared and unsheared regions, is within the near-wall electric double
layer. Thus the uniform velocity, which located in the unsheared
region, is dominating the velocity profile. The uniform velocity
arises due to a yield stress and has similarity with the classical
plug-like velocity of EOF of a fluid without a yield stress. It is
important to mention that the dominant velocity (uniform velocity)
here arises because of the result of interaction between rheological
properties of fluid and electric forcing. Thus it is subject to more
controlling factors than the classical plug-like velocity of EOF of
non-yield stress fluid.

Again, since the effect of yield-stress strongly sensitive on the
electric forcing, another concern of this study is to analyze EOF of a
yield-stress fluid for the situation when the zeta-potential can be
large or small. Under the consideration of small electric potential,
we may get the Debye-Huckel approximation and it is often applied to
obtain the Poisson-Boltzmann equation in linear form. However, this
model can cope with both the limits of small and large zeta potentials
without the Debye-Huckel approximation. Zeta potential is an essential
factor in the model since both shear stress distribution and thus the
velocity profile throughout the entire tube requires the zeta
potential as a boundary condition. Although, EOF of non-Newtonian
fluids under a large zeta potential has been carried out by many
researchers, e.g., Bharti et al. \cite{Bharti}, Zhao and Yang
\cite{Zhao2,Zhao3}, Vasu and De \cite{Vasu}, Babaie et
al. \cite{Babaie} and Vakili et al. \cite{Vakili}; they have not
included yield stress in their studies.

From the various types of non-Newtonian models, the power-law model
has been chosen mostly to study the electrokinetic flow of
non-Newtonian fluids analytically or numerically
\cite{Berli1,Berli2,Olivares,Bharti,Zhao2,Vasu,Babaie,Chakraborty,Zhao4,Tang2,Zhao5}. However,
the nonlinear interaction between the hydrodynamic and electric
forcings in driving a power-law or more generally a Herschel-Bulkley
fluid through a non-uniform vessels has not been taken into account. It
may be noted that it is not generally possible to solve the nonlinear
rheological problem of non-Newtonian fluid flow
analytically. Nevertheless, most of the analytical researchers
\cite{Berli1,Das,Zhao4,Zhao6,Ng2} on EOF of a power-law fluid have
considered simple geometries (i.e. flow over a flat surface, or
through a uniform parallel-plate or circular channel). 

It is aimed to study pressure-driven and electroosmotic flow of a
Herschel-Bulkley fluid in a non-uniform micro-vessels. The study deals
with the analytical analysis concerning how the hydrodynamic and
electric forcings will interact with each other in driving
non-parallel flow of a Herschel-Bulkley fluid through a non-uniform
vessel. The study has been analyzed with the help of the lubrication
theory \cite{Ajdari1,Ajdari2,Long,Ghosal,Ng3,Ng4} together with taking into
account the near-wall depletion effect. Therefore, the study is valid
particularly for the cases where the Reynolds number is small and the
ratio between length scales in the radial direction and the axial
direction of the vessel is much smaller than unity. The radius of the
vessel as well as the wall charge has been considered to change along
with the radius of the tube. It is to be investigated whether for
non-parallel flow, the superposition of components due to the two
forcings does work linearly for a non-Newtonian fluid, even if we
consider a Newtonian depletion layer. Following MacInnes et
al. \cite{MacInnes}, Ng and Qi \cite{Ng5}; the EDL is not resolved
directly for the electrokinetic transport and the
Helmholtz–Smoluchowski (HS) slip boundary condition
\cite{Zimmerman,Ng5} has been considered here instead.

We have considered the radius of the vessel of the order of 100 $\mu
m$. It is much larger than the depletion layer thickness ($100 nm$)
which is bigger than the EDL thickness ($10 nm$). The analysis has
been simplified focusing on the bulk fluid flow only (without the
direct involvement with the depletion layer). Berli and Olivares
\cite{Berli1} used this ‘‘single fluid’’ approach in order to study
the hydrodynamic and EOF with slip at the wall as a limiting case.

The current investigation can be applied to model the flows of various
bio-fluids in far ranging technological implications, which
predominantly exhibit power-law/Herschel-Bulkley fluid (e.g. blood,
bio-fluids, protein solutions, DNA solutions, colloidal suspensions,
solutions of high polymers and suspensions
\cite{Berli1,Misra2,Barnes,Das,Tang,Ng,Bharti,Vakili,Chakraborty,Zhao4,Ng2,Ng5,Dey,Ranjit,Ranjit2,Mondal,Ranjit3,Ranjit4})
behavior and the dynamics of which are not addressed properly. The
results can also be utilized as a potential platform to build up
microfluidic Lab-on-a-chip based devices dealing with the
power-law/Herschel-Bulkley type bio-fluids, with electrokinetically
driven flow actuation mechanisms. Nonlinear effects, which are
expected when complex fluids are subjected to electric potential
and/or pressure gradients in micro-channels, are vital aspect for the
design and operation of microfluidic chips. Thus the interactions
between fluid rheology, slip wall velocity, and EDL phenomena may be
exploited to achieve a surprising augmentations in the device
throughput over characteristic regimes of interest.

\section{Formulation}
Pressure-driven and Electroosmotic flow (EOF) of a non-Newtonian fluid
is intended to analyze theoretically in an axisymmetric cylindrical
micro-vessel. The fluid is undertaken as an incompressible viscous
Herschel-Bulkley fluid under the consideration that the radius and the
wall potential may vary slowly and periodically along the axis of the
vessel. In addition, the diverging/tapered nature of the mean tube
radius has also been incorporated in this study.

\begin{figure}
\begin{center}
\psfrag{Wall shape: R=h(Z)}{{\scriptsize Wall shape: $R=H(Z)$}}
\includegraphics[width=3.in,height=2.0in]{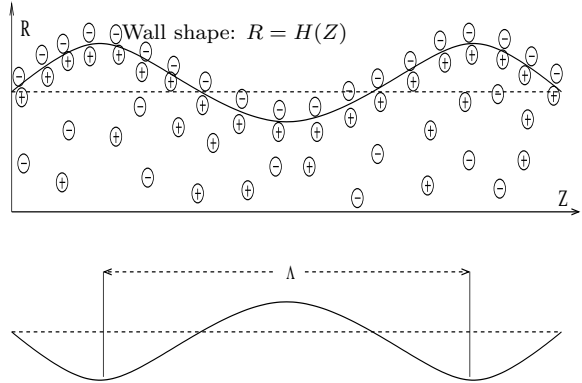}
\caption{A physical sketch of the electric double layer (EDL) next to
  a negatively charged solid surface of the problem for transverse
  pressure-driven and electroosmotic flow of a Herschel-Bulkley fluid
  through a non-uniform tube with undulated walls and charge-modulated
  surfaces.}
\label{manuscript_geofig1}
\end{center}
\end{figure}

Let us assume (R,$\theta$,Z) as the cylindrical polar coordinates of
the location of any fluid particle (cf. Fig. \ref{manuscript_geofig1})
in the micro-vessel. $R$ is measured along the radius of the vessel, Z
is being measured along the axis of the vessel and $\theta$ is
considered as the rotational coordinate. $R=H (Z)$
(cf. Fig. \ref{manuscript_geofig1}) represents the wall of the
micro-vessel and the wall potential is taken as $\xi=\xi (Z)$. Both
$H (Z)$ and $\xi (Z)$ have been treated as periodic functions of $Z$
with the same wave length $\lambda$. The radius of the vessel is
considered as much smaller than $\lambda$ so that ratio between them
to be much smaller than unity. In order to apply lubrication theory,
the Reynolds number of the flow is considered to be small together
with small ratio of the tube radius and wavelength as said above.

The EDL is a very thin region close to the charged surface where an
excess of counterions over coions in order to neutralize the surface
charge \cite{Probstein}. Thus, a net fluid flow can be driven in the
desired direction when imposing external electric field
\cite{Park}. The application of a time-periodic external electric
field creates a net body force on the free ions inducing a
time-periodic bulk motion within the electric double layer (EDL)
\cite{Khan}. (governing equation with electrical body force)

The non-Newtonian bulk fluid has been treated as a viscous
Herschel-Bulkley fluid. It has been considered that the walls are
non-adsorbing indicating that the layer very near the boundary is a
depletion layer. In this layer the fluid, owing to the absence of the
inclusion of the macromolecules, is much less viscous than the bulk
fluid and we may treat it as Newtonian. We further consider that the
electric double layer is thinner than the depletion layer. As a
result, the electrokinetic effect can be confined to the region of
Newtonian fluid near the boundary of the vessel. Moreover, the
depletion layer thickness is considered much smaller than the radius
of the tube. Thus, the electrokinetic forcing has been simplified to a
Helmholtz-Smoluchowski (HS) Electroosmotic slip velocity
\cite{Zimmerman}. As mentioned by Dutta and Beskok \cite{Dutta}, the
HS electroosmotic velocity can be generated as a result from the
balance between the viscous diffusion terms and the electroosmotic
forces in the absence of the pressure gradients.

However, the employed simplifying assumption on HS slip velocity can
be reasonable in cases with very thin EDL or in flow fields with not
strong non-Newtonian rheological behavior. It is important to mention
that there are a lot of studies in the literature which provide
equations to modify the HS velocity for non-Newtonian fluid flows.

EOF of rheological fluid has been considered to be governed by the
Navier-Stokes equations
\begin{eqnarray}
\nabla \cdot{\bf V}=0
\end{eqnarray}
\begin{eqnarray}
\rm{and~~~~}\rho \frac{d{\bf V}}{dt}=\nabla\sigma+ \rho_e {\bf E}.
\end{eqnarray}
Here ${\bf V}$ represents the velocity vector, ${\bf E}$ is the
electrical field strength applied to the micro-vessel, $\rho$ being the
fluid density, $\rho_e$ is the local net charge density and
$\frac{d}{dt}$ is the material time-derivative. $\sigma$ defines the
Cauchy stress tensor which can be expressed as
\\ $~~~~~~~~~~~~~~~~~~~~~~~~~~~~~~~~~~~~~~~~~~~~~~~~~~~~~\sigma=-PI+T$,\\in
which\\$~~~~~~~~~~~~~~~~~~~~~~~~~~~~~~~~~~~~~~~~~~~~~~~~~~~~~T=2\mu
E_{ij}+\eta IS $ and $S=\nabla\cdot {\bf V}$,\\ $E_{ij}$ being the
symmetric part of the velocity gradient, is defined
by\\$~~~~~~~~~~~~~~~~~~~~~~~~~~~~~~~~~~~~~~~~~~~~~~~~~~~~E_{ij}=\frac{1}{2}[L+L^T]$,\\$~~~~~~~~~~~~~~~~~~~~~~~~~~~~~~~~~~~~~~~~~~~~~~~~\rm{where}~~L=\nabla{\bf
  V}$. \\ $-PI$ is known as the indeterminate part of the stress due
to the constraint of incompressibility. We denote $\mu$ and $\eta$ as
viscosity parameters.

As mentioned above, Herschel-Bulkley model represents the combined
effect of Bingham plastic and power-law behavior of the fluid. It is
well known that when strain-rate $\dot{\gamma}$ is less than
$\frac{\tau_0}{\mu_0}$, the fluid behaves like a viscous fluid with
constant viscosity $\mu_0$. But as the strain rate increases and the
yield stress threshold ($\tau_0$) is reached, the fluid behavior is
better described by a power law of the form
\begin{eqnarray}
\tau_{ij}=\left\{\begin{array}{r@{\quad : \quad}l}(\mu\dot{\gamma}^{n-1}+\frac{\tau_0}{\dot{\gamma}})\dot{\gamma_{ij}}~ & for~\tau\ge\tau_0,
\\\dot{\gamma_{ij}}=0~ & for~\tau<\tau_0
\end{array} \right.,\\
\end{eqnarray}
in which
\begin{eqnarray}\dot{\gamma_{ij}}=\left(\frac{\partial U_i}{\partial
  X_j}+\frac{\partial U_j}{\partial X_i}\right),~~\tau=\sqrt{\frac{\tau_{ij}\tau_{ij}}{2}},~~\dot{\gamma}=\sqrt{\frac{\dot{\gamma_{ij}}\dot{\gamma_{ij}}}{2}}.
\end{eqnarray}
Here $\mu$ and n denote respectively the consistency factor and the
power law index. Where $n<1$ and $n>1$ correspond to a shear thinning fluid
and shear thickening fluid respectively.

Lee et al. \cite{Lee} reported that the driving force for the
electroosmotic flow, $\rho_e {\bf E}$, appears only in EDL as the
local net charge density is non-zero only in the EDL. The EOF velocity
varies sharply in a thin layer of liquid near the wall from zero at
the channel wall surface to an approximately constant value at the
outer edge of the EDL \cite{Lee}.

The general micro-channel height deviates from 10 to 100 $\mu m$ and
thickness of EDL changes from 1 to 10 $nm$ in a typical microfluidic
device \cite{Kim}. Buffer solutions have a concentration of the order
of $mM$ for most on-chip microfluidic applications resulting in a very
thin EDL \cite{Lee,Khan}. As a example, a solution with a 10 $mM$
concentration has a EDL thickness of approximately 10 $nm$ that is
negligible in comparison with the micro-channel diameter (e.g. 100 $\mu
m$) \cite{Lee}.

Hence, the driving force term $\rho_e {\bf E}$ will be dropped off in
the governing equation and the bulk fluid flow, outside of the EDL
region, can be modeled by replacing the electroosmotic body force with
help of introduction of a HS slip boundary condition at the channel
wall \cite{Lee,Khan}. Moreover, as reported by Kim et al. \cite{Kim},
$\rho {\bf E}$, can be dropped off and the electroosmotic effect can
be replaced by introducing a slip velocity at the channel wall for the
modeling purpose of the bulk fluid flow outside the EDL following HS
formulation on the basis of the assumption that both viscosity and
permittivity are constant and they are independent of time and space
\cite{Kim}.

 Although the HS formulation is originally introduced for
steady state DC EOF, it can be employed for low frequency AC (or time
periodic) EOF since the charge relaxation frequency ($f_c =
(1/2\pi)(\sigma/\epsilon)\cong 10^6 \sim 10^8 HZ$ be much larger for
equilibrium distribution of ions in the electrical double layer
\cite{Green}. Furthermore, HS formulation is widely employed for
non-uniform electroosmotic flow model in capillaries \cite{Potocek}
and micro-channels \cite{Horiuchi,Lee,Chang}.

This slip boundary condition can also be used for time-periodic EOF
since the applied electric field frequency ($10^2\sim 10^5~ Hz$ ) is
less than the charge relaxation frequency ($10^6\sim 10^8~Hz$)
\cite{Green2} although it was originally introduced for steady-state
EOF. Furthermore, the Helmholtz-Smoluchowski development-based slip
boundary condition is widely employed for non-homogeneous surface
charges and time-periodic EOF analysis
\cite{Kim,Lee,Potocek,Chang,Ajdari1}. Since the Debye length of the
electric double layer is less than 10 $nm$ in the micro-channels used on
a chip for many laboratory applications \cite{Zimmerman}, inside such
a thin region it is hardly available any polymer molecules due to the
depletion effect \cite{Tuinier} indicating that the solvent viscosity
to be the appropriate viscosity for determination of the boundary HS
slip velocity \cite{Zimmerman}.

Thus replacing the electroosmotic body force (i.e. $\rho_e {\bf E}$)
with help of introduction of a HS slip boundary condition, the
governing equations (Navier-Stokes equations) of the Herschel-Bulkley
fluid may be written in the form

\begin{equation}
\rho \left (\frac{\partial U}{\partial t}+U\frac{\partial
U}{\partial Z}+V\frac{\partial U}{\partial R}\right
)=-\frac{\partial P}{\partial Z}+\frac{1}{R}\frac{\partial
  (R\tau_{RZ})}{\partial R}+\frac{\partial \tau_{ZZ}}{\partial Z},
\end{equation}
\begin{equation}
\rho\left (\frac{\partial V}{\partial t}+U\frac{\partial V}{\partial
Z}+V\frac{\partial V}{\partial R}\right )=-\frac{\partial
P}{\partial R}+\frac{1}{R}\frac{\partial
  (R\tau_{RR})}{\partial R}+\frac{\partial \tau_{RZ}}{\partial Z}.
\end{equation}

The following non-dimensional variables will be introduced in the
analysis:
\begin{eqnarray}
\bar{Z}=\frac{Z}{\lambda},~~\bar{R}=\frac{R}{a_0},~~\bar{U}=\frac{U}{c},~\bar{V}=\frac{V}{c\delta},~
\delta=\frac{a_0}{\lambda},~\bar{P}=\frac{a_0^{n+1}P}{\mu
  c^n\lambda},~\bar{t}=\frac{ct}{\lambda},
~h=\frac{H}{a_0},~\phi=\frac{d}{a_0},\nonumber\\ Re=\frac{\rho
  a_0^n}{\mu
  c^{n-2}},~\bar{\tau}_0=\frac{\tau_0}{\mu\left(\frac{c}{a_0}\right)^n},~\bar{\tau}_{RZ}=\frac{\tau_{RZ}}{\mu\left(\frac{c}{a_0}\right)^n},~\bar{\tau}_{ZZ}=\frac{\tau_{ZZ}}{\mu\delta\left(\frac{c}{a_0}\right)^n},~\bar{\tau}_{RR}=\frac{\tau_{RR}}{\mu\delta\left(\frac{c}{a_0}\right)^n},~~~~~~~~
\label{manuscript_non-dimensionalize}
\end{eqnarray}
where we consider $\lambda$, $a_0$ as the length scales for velocity
variations in the $x,y$-direction, respectively and $c$ as a scale for
the velocity in the x-direction. The equation governing the flow of
the fluid can now be rewritten in the form (dropping the bars over the
symbols):
\begin{equation}
 Re\delta \left (\frac{\partial U}{\partial t}+U\frac{\partial
U}{\partial Z}+V\frac{\partial U}{\partial R}\right
)=-\frac{\partial P}{\partial Z}+\frac{1}{R}\frac{\partial
  \left(\Phi\left(R\frac{\partial U}{\partial R}+R\delta^2\frac{\partial V}{\partial
Z}\right)\right)}{\partial R}+2\delta^2\frac{\partial \left(\Phi\frac{\partial
U}{\partial Z}\right)}{\partial Z},
\end{equation}
\begin{equation}
Re\delta^3\left (\frac{\partial V}{\partial t}+U\frac{\partial V}{\partial
Z}+V\frac{\partial V}{\partial R}\right )=-\frac{\partial
P}{\partial R}+\delta^2\frac{2}{R}\frac{\partial
  \left(R\Phi\frac{\partial V}{\partial R}\right)}{\partial R}+\delta^2\frac{\partial\left(\Phi\left(\frac{\partial U}{\partial R}+\delta^2\frac{\partial V}{\partial
Z}\right)\right)}{\partial Z}
\end{equation}
\begin{eqnarray},
where~~~\Phi=\left|\sqrt{2\delta^2\left\{\left(\frac{\partial V}{\partial R}\right)^2+\left(\frac{V}{R}\right)^2+\left(\frac{\partial
U}{\partial Z}\right)^2\right\}+\left(\frac{\partial U}{\partial R}+\delta^2\frac{\partial V}{\partial
Z}\right)^2}\right|^{n-1}\nonumber\\+\tau_0\left|\sqrt{2\delta^2\left\{\left(\frac{\partial V}{\partial R}\right)^2+\left(\frac{V}{R}\right)^2+\left(\frac{\partial
U}{\partial Z}\right)^2\right\}+\left(\frac{\partial U}{\partial R}+\delta^2\frac{\partial V}{\partial
Z}\right)^2}\right|^{-1}
\end{eqnarray}
For the case $\delta\ll 1$ and $\delta Re\ll 1$ (i.e. the lubrication approximation), the governing
equations, describing the flow in terms of the dimensionless variables
(\ref{manuscript_non-dimensionalize}), assume the form
\begin{equation}
0=-\frac{\partial P}{\partial Z}+ \frac{1}{R}\frac{\partial
  (R\frac{\partial U}{\partial R}|\frac{\partial U}{\partial
    R}|^{n-1}+R\tau_0)}{\partial R}
\label{manuscript_Z_momentum_lubrication}
\end{equation}
\begin{equation}
\rm{and~~~}0=-\frac{\partial P}{\partial R}.
\end{equation}
The boundary conditions \cite{Dutta,Zimmerman,Probstein,Hunter,Kim,Rubin,Khan} in non-dimensional
co-ordinates take the form
 \begin{eqnarray}
 \frac{\partial U}{\partial R}=0,~ \tau_{RZ}=0~ at~ R=0~~~~~~~~~~~~~~~~~~~~~~~\rm{and~~~}
\label{manuscript_non-dimensional_boundary-condition_central}\\U=U_s(Z)=-\frac{\zeta E_Z}{\mu_s}\xi(Z)~ \rm{at}~ R=h(Z),~~~~~~~~~~~~~~~~~~~~~~~~~~~~~~~
\label{manuscript_non-dimensional_boundary-condition_wall}
\end{eqnarray}
where $\zeta$ $\&$ $\mu_s$ denote the permittivity and
dynamic viscosity of the solvent in the depletion layer respectively and $E_Z$ is
the applied axial electric field.

Applying the condition
(\ref{manuscript_non-dimensional_boundary-condition_central}),
the solution of equation
(\ref{manuscript_Z_momentum_lubrication}) is found as
\begin{equation}
\frac{\partial U}{\partial R}\left|\frac{\partial U}{\partial
    R}\right|^{n-1}+\tau_0=\frac{R}{2}\frac{\partial P}{\partial Z}.
\label{manuscript_axial_velocity_1st_derivative}
\end{equation}

Now, using the condition
(\ref{manuscript_non-dimensional_boundary-condition_wall}), the
solution of equation (\ref{manuscript_axial_velocity_1st_derivative})
can be expressed as
\begin{equation}
U(R,Z,t)=U_s(Z)+\frac{1}{(m+1)P_1}\left[(P_1h-\tau_0)^{m+1}-(P_1R-\tau_0)^{m+1}\right],~
0\le R\le h(Z);
\label{manuscript_axial_velocity}
\end{equation}
where~$P_1=-\frac{1}{2}\frac{\partial P}{\partial Z}
\rm{~~and~}~m=\frac{1}{n}$.

If we consider $\tau_0=0$ (i.e. the absence of yeild stress) and $n=1$
(i.e. for Newtonian fluid), the equation
(\ref{manuscript_axial_velocity}) can be simplified as
\begin{equation}
U(R,Z,t)=U_s(Z)+\frac{P_1}{2}\left[h^{2}-R^{2}\right],~
0\le R\le h(Z).
\label{manuscript_axial_velocity_siplified}
\end{equation}
Thus, the results are found to coincide with those obtained by Dutta
and Beskok \cite{Dutta} (Ref. equation (13) and (14) of \cite{Dutta}),
when the flow in two-dimensional channel is replaced by the flow in
axisymmetric tube. This observation may be considered as a validation
of the present study.

The radius of the plug flow region, $R_0$
is given by
\begin{eqnarray*}
\frac{\partial U}{\partial R} =0~~at~R=R_0.
\end{eqnarray*}
Which in view of (\ref{manuscript_axial_velocity}), is obtained as
\begin{eqnarray*}
R_0=\tau_0/P_1.
\end{eqnarray*}
Considering $\tau_{RZ}=\tau_h$ at R=h, we find $h=\tau_h/P_1$ and
\begin{eqnarray}
\frac{R_0}{h}=\frac{\tau_0}{\tau_h}=\nu~(say),~~0<\nu<1.~~~~~~~~~~~~
\end{eqnarray}
We have then obtained the expression of the plug velocity as
\begin{eqnarray}
U_P=\frac{(P_1h-\tau_0)^{m+1}}{(m+1)P_1}.~~~~~~~~~~~~~~~~~~~~~~~~~~~~~
\label{manuscript_axial_plug_velocity}
\end{eqnarray}

The instantaneous rate of volume flow through the micro-vessel, $Q$, may then be given by
\begin{eqnarray}
Q=2\int_{0}^{R_0}RUdR+2\int_{R_0}^{h}RUdR~~~~~~~~~~~~~~~~~~~~~~~~~~~~~~~~~~~~~~~~~~~~~~~~~~~~~~~~~~~\nonumber~~~\\=U_s(Z)h^2+\frac{P_1^m(h-R_0)^{m+1}(h^2(m^2+3m+2)+2R_0h(m+2)+2R_0^2)}{(m+1)(m+2)(m+3)},~~m=\frac{1}{n}.
\label{manuscript_volume_flow_rate}
\end{eqnarray}
For $P_1< 0$ (i.e., under adverse pressure gradient), the stress is
positive. From equation (\ref{manuscript_volume_flow_rate}), it is clear
that $P_1< 0$ where $Q< U_s(Z)h^2$ and $P_1> 0$ where $Q>
U_s(Z)h^2$. A single expression for $P_1$, the pressure gradient, as a
function of the flow rate, is then obtained as
\begin{eqnarray}
P_1=-\frac{1}{2}\frac{\partial P}{\partial
  Z}=\left[\frac{(m+1)(m+2)(m+3)}{(h-R_0)^{m+1}(h^2(m^2+3m+2)+2R_0h(m+2)+2R_0^2)}\right]^n|Q-U_s(Z)h^2|^{n-1}(Q-U_s(Z)h^2)\nonumber~~~~\\=\left[\frac{(m+1)(m+2)(m+3)}{h^{m+3}(1-\nu)^{m+1}((m^2+3m+2)+2\nu
    (m+1)+2\nu^2)}\right]^n|Q-U_s(Z)h^2|^{n-1}(Q-U_s(Z)h^2).~~~~~~~~~~~~~~~~~~~
\label{manuscript_pressure_gradient}
\end{eqnarray}
The net pressure drop across one wavelength may now be evaluated as
\begin{eqnarray}
 \Delta P=\int_{0}^{1}\left(\frac{\partial P}{\partial
   Z}\right)dZ~.
\label{manuscript_pressure_difference_wave_length}
\end{eqnarray}
For the Newtonian fluid ($n=1$), the solution of the equation
(\ref{manuscript_pressure_difference_wave_length}) is given by 

\begin{eqnarray}
Q_{n=1}=(\Delta P-I_2)/I_1,~~~~~~~~~~~~~~~~~~~~~~~~~~~~~\\
\rm{where}~~I_1=-24\int_{0}^{1}\frac{1}{(h-R_0)^{2}(3h^2+3R_0h+R_0^2)}dZ,\nonumber\\I_2=-24\int_{0}^{1}\frac{U_s(Z)h^2}{(h-R_0)^{2}(3h^2+3R_0h+R_0^2)}dZ.\nonumber
\label{manuscript_time_average_flow_flux_relation}
\end{eqnarray}

For the uniform vessel, a special case where $h$ and $U_s(Z)$ are both
constants, equation (\ref{manuscript_pressure_difference_wave_length})
may also yield

\begin{eqnarray}
Q_{uniform}=\mp\frac{(h-R_0)^{m+1}(h^2(m^2+3m+2)+2R_0h(m+2)+2R_0^2)}{(m+1)(m+2)(m+3)}|\Delta
P|^m+U_s(Z)h^2,
\label{manuscript_time_average_flow_flux_relation_uniform_tube}
\end{eqnarray}
where the upper or lower signs are for $\Delta P>0$ or $\Delta P<0$ respectively.

In general, the equation
(\ref{manuscript_pressure_difference_wave_length}) cannot be
integrated in closed form. For further investigation of the problem
under consideration, it may be solved numerically under an efficient
iterative solution scheme. The left side of equation
(\ref{manuscript_pressure_difference_wave_length}) has been calculated
based on the $i$th trial flow rate $Q^{(i)}$ in order to get the
$i$th trial pressure drop $\Delta P^{(i)}$. It is then compared with
the given $\Delta P$. The following equations show how the $(i+1)$th
flow rate will be found by the difference of the two:
\begin{eqnarray}
Q^{(i+1)}=Q^{(i)}+\frac{\Delta P-\Delta P^{(i)}}{d\Delta
  P/dQ},
~~~~~~~~~~~~~~~~~~~~~~~~~~~~~~~~~~~~~~~~~~~~~~~~~~~~~~~~~~~~~~~\\ \rm{where}~~
\frac{d\Delta
  P}{dQ}=n\left\{\frac{(m+1)(m+2)(m+3)}{(m+1)(m+2)+2(m+1)\nu+2\nu^2)(1-\nu)^{(k+1)}}\right\}^n\int_{0}^{1}\frac{|Q-U_s(Z)h^2|^{(n-1)}}{h^{(3n+1)}}dZ.~~~~
\label{manuscript_time_average_flow_flux_iteration}
\end{eqnarray}
We have taken the Newtonian flow rate $Q_{n=1}$ given by equation
(\ref{manuscript_time_average_flow_flux_relation}) as the initial
value of $Q^{(1)}$. It is normally observed that the solution
converges in 5 iterations. However, the accuracy of the solution
  depends on the numerical evaluation of the definite integrals and
  time required for these solutions is much more when the tube radius
  is of converging/diverging nature.
\begin{figure}
\centering
\includegraphics[width=3.5in,height=2.0in]{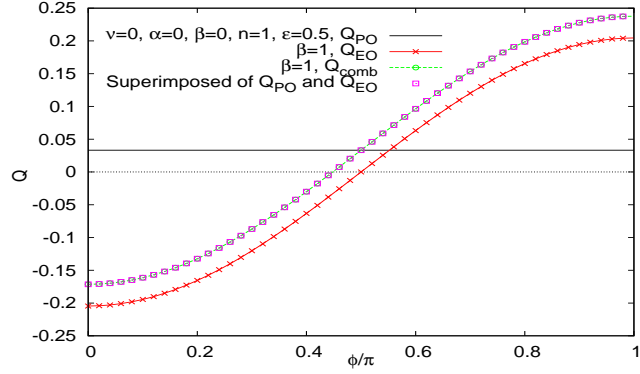}
\caption{Variation of flow rate $Q$ with the phase shift $\phi$ when
  $k=0;~\Delta P=0,~-1;~\nu=0;~\epsilon=0.5;~\alpha=0;~\beta=0,~1$ for a
  Newtonian fluid.}
\label{manuscript_flow_rate_phase_diff_Newt2.1}
\end{figure}

\begin{figure}
\includegraphics[width=3.5in,height=2.0in]{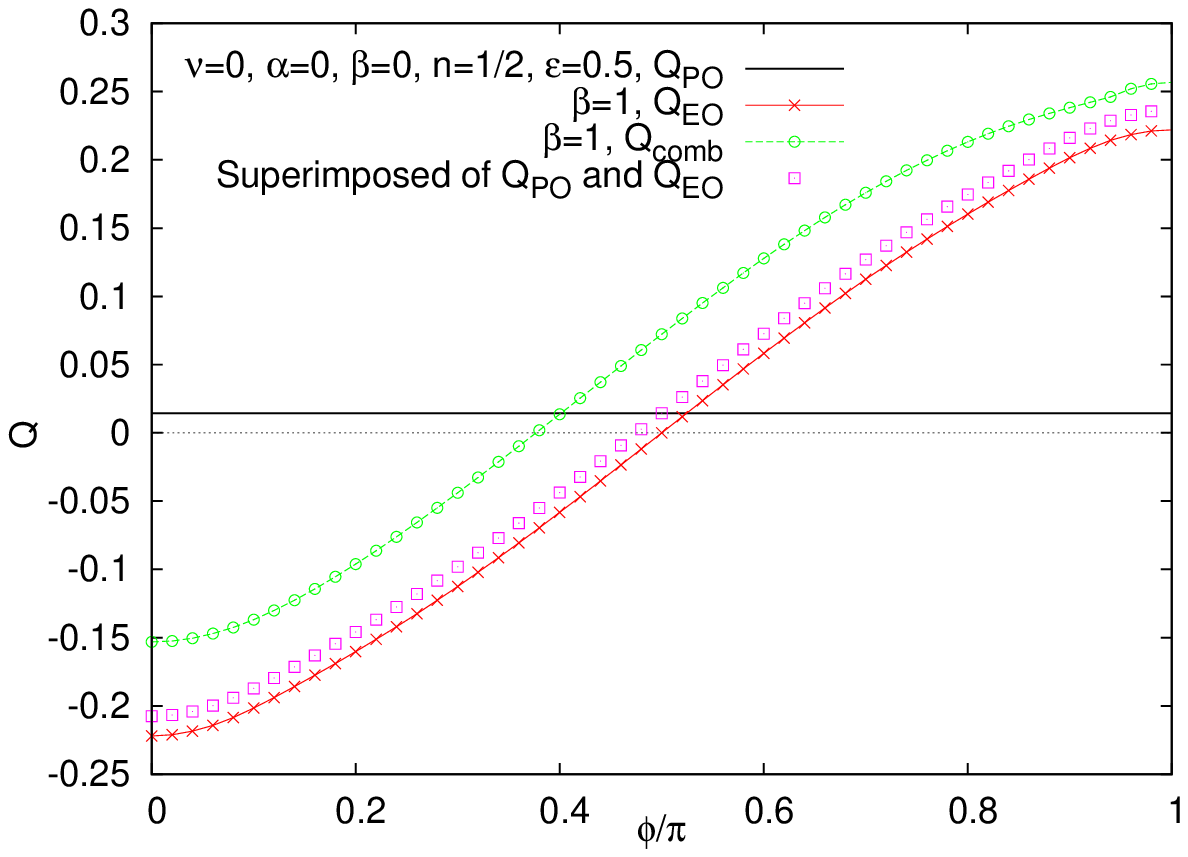}\includegraphics[width=3.5in,height=2.0in]{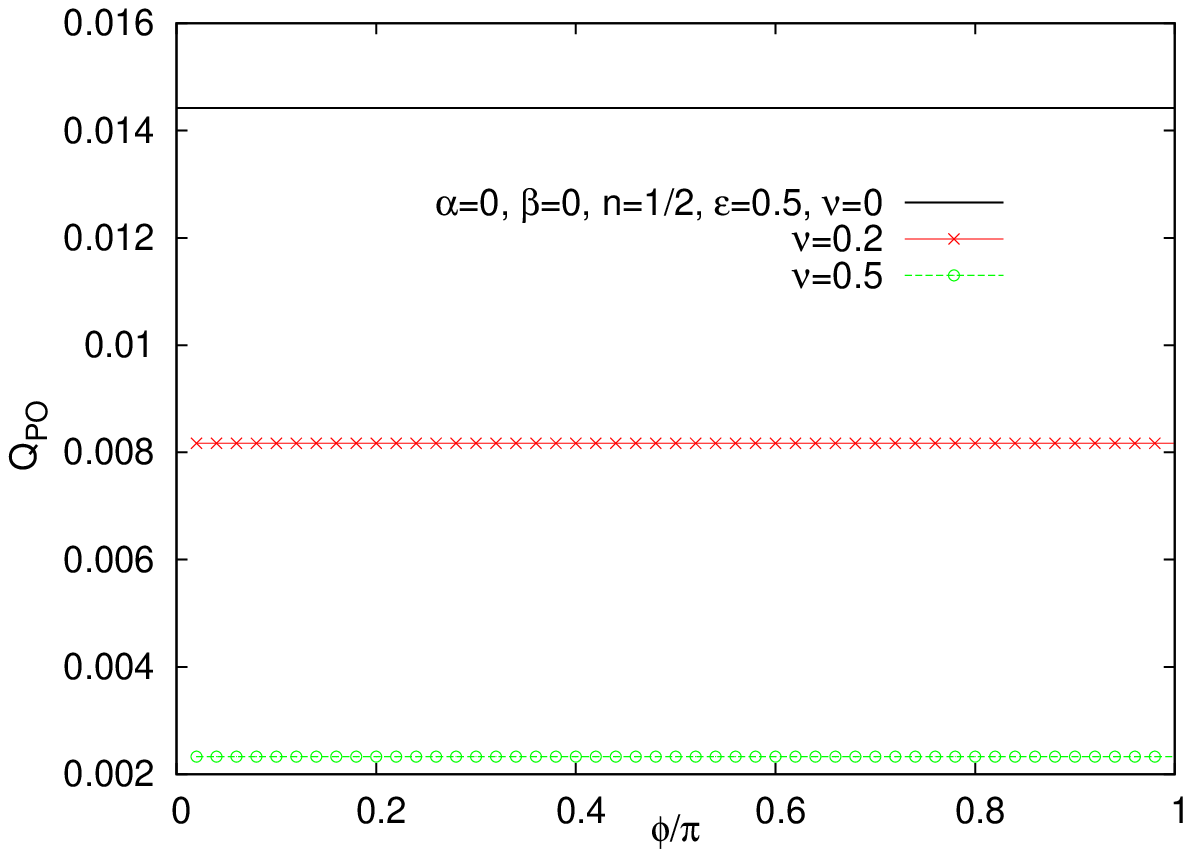}
\\$~~~~~~~~~~~~~~~~~(a)~~~~~~~~~~~~~~~~~~~~~~~~~~~~~~~~~~~~~~~~~~~~~~~~~~~~~~~~~~~~~~~~~~~(b)~~~~~~~~~$
\includegraphics[width=3.5in,height=2.0in]{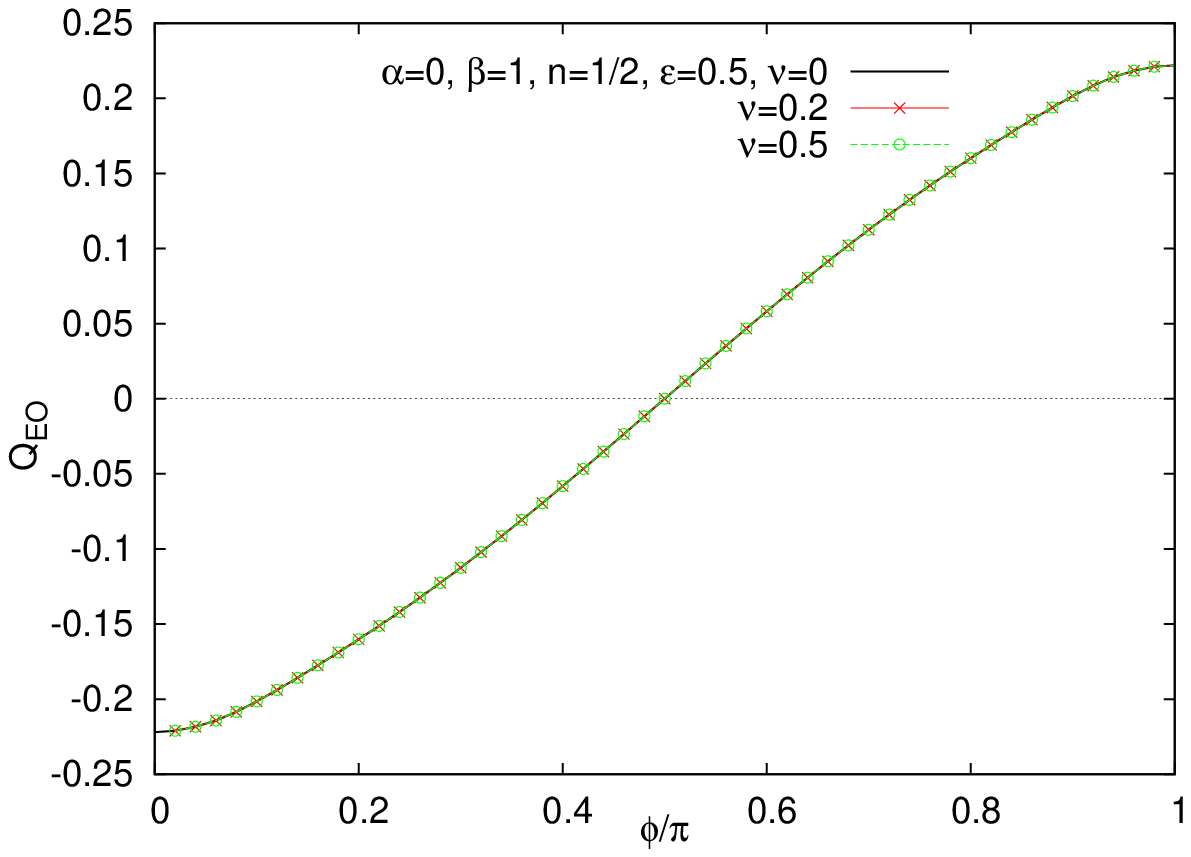}\includegraphics[width=3.5in,height=2.0in]{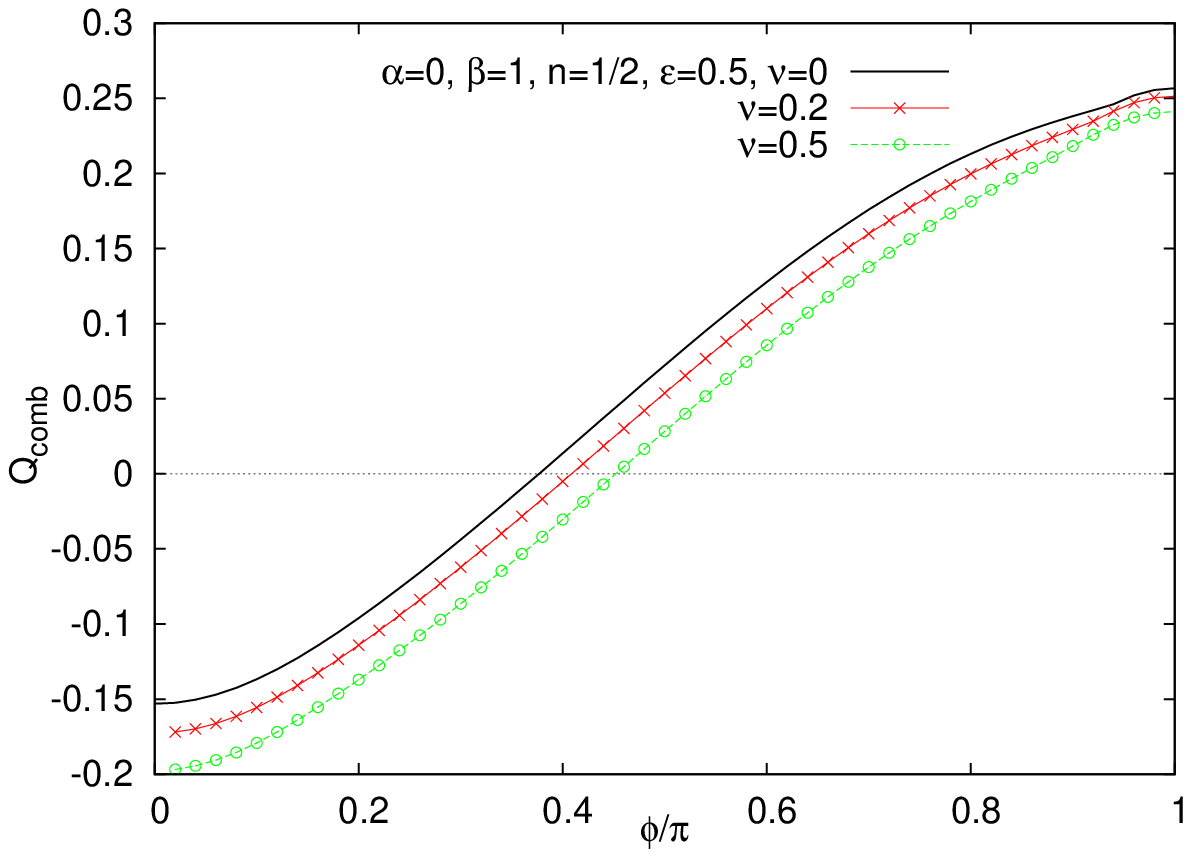}
\\$~~~~~~~~~~~~~~~~~(c)~~~~~~~~~~~~~~~~~~~~~~~~~~~~~~~~~~~~~~~~~~~~~~~~~~~~~~~~~~~~~~~~~~~(d)~~~~~~~~~$
\caption{Variation of flow rate $Q$ with the phase shift $\phi$ when
  $k=0;~\Delta P=0,-1;~\nu=0~\text{to }0.5;~\epsilon=0.5;~\alpha=0;~\beta=0,~1$ for a
  shear-thinning fluid.}
\label{manuscript_flow_rate_phase_diff_shear_thin1.1-1.4}
\end{figure}

\section{Quantitative Investigation}
In this section, our aim is to study the problem and present the
computational results for the said quantities
graphically. Computations of the quantities have been carried out by
extensive use of the software Mathematica. In order to get the
computational results, 8 input, i.e. $n$, $\Delta P$, $\epsilon$,
$\alpha$, $\beta$, $\phi$, $k$ and $\nu$ have been used. As long as $h$ and $U_s$ are
considered as slowly varying functions of Z, equations
(\ref{manuscript_axial_velocity}),
(\ref{manuscript_pressure_gradient})-(\ref{manuscript_time_average_flow_flux_relation})
are valid for them. The radius of the tube and HS slip velocity have
been treated as the following sinusoidal functions
\begin{eqnarray}
h=1+kZ+\epsilon \cos (2\pi Z),~~~U_s=\alpha+\beta \cos (2\pi Z+\phi);
\label{manuscript_h_Us}
\end{eqnarray}
where $\alpha$ stands for the mean HS slip velocity, $\epsilon$ being
the amplitudes of the wall shape, $\beta$ for slip modulation and
$\phi$ for the phase shift between the two modulation waves. Here, k
is a constant parameter whose positive value means that the mean tube
radius is of diverging nature, while the negative value of it
indicates the tapered nature of the mean tube radius. The above
idealized wave forms of the geometric and electrokinetic modulation
may provide indicative results for the physics involved in this
problem.

It is easy to note that, in either of the two particular cases,
e.g. Newtonian limit or uniform tube radius, there is no essential
coupling between the two forces hydrodynamic and electrokinetic. Hence
the flow is a linear combination of the hydrodynamic and
electrokinetic effects in these cases. For another special case where
$\Delta P=0$ and walls are flat ($\epsilon =0$), it is worthwhile to
note that
\begin{eqnarray}
Q=\alpha~~ \rm{for~ any}~~ n~~ \rm{and}~~ \beta. 
\label{manuscript_Q_uniform_wall_fre_pump}
\end{eqnarray}
In the numerical discussion, these limiting and special cases will be
further discussed.

In order to inspect the issue of linear superposition of the
hydrodynamic and electrokinetic effects closely, we define the following types
of flow rate $Q$:
\begin{eqnarray}
Q_{PO}=\rm{rate~ of~ flow~ due~ to~ hydrodynamic~ forcing}~~\Delta P~~\rm{only},\\
Q_{EO}=\rm{rate~ of~ flow~ due~ to~ electric~ forcing}~~U_s~~\rm{only},~~~~~~~~~~\\
Q_{comb}=\rm{rate~ of~ flow~ due~ to~ combined~ action~ of}~~ \Delta P \rm{and}~~U_s.
\end{eqnarray}

\begin{figure}
\includegraphics[width=3.5in,height=2.0in]{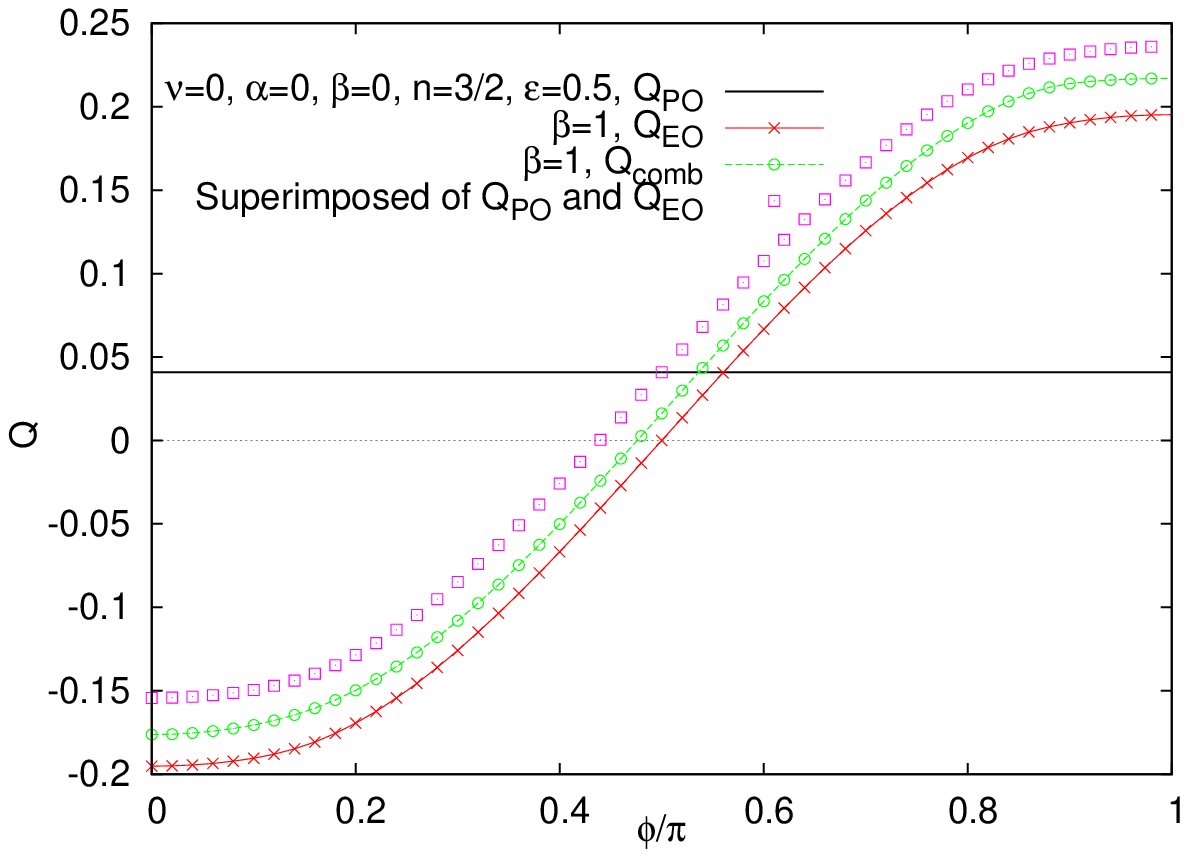}\includegraphics[width=3.5in,height=2.0in]{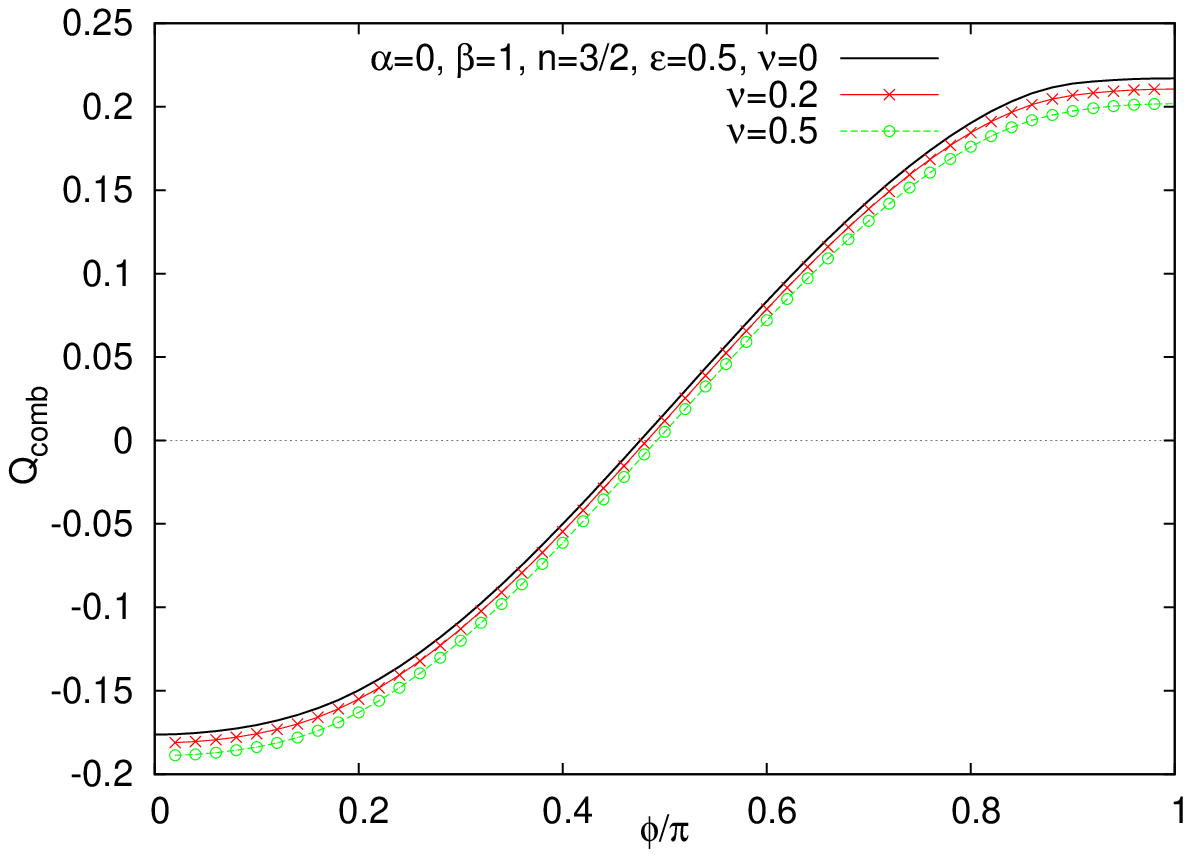}
\\$~~~~~~~~~~~~~~~~~(a)~~~~~~~~~~~~~~~~~~~~~~~~~~~~~~~~~~~~~~~~~~~~~~~~~~~~~~~~~~~~~~~~~~~(b)~~~~~~~~~$
\caption{Variation of flow rate $Q$ with the phase shift $\phi$ when
  $k=0;~\Delta P=0,~-1;~\nu=0~\text{to }0.5;~\epsilon=0.5;~\alpha=0;~\beta=0,~1$ for a
  shear-thickening fluid.}
\label{manuscript_flow_rate_phase_diff_shear_thick3.1-3.4}
\end{figure}

\begin{figure}
\includegraphics[width=3.5in,height=2.0in]{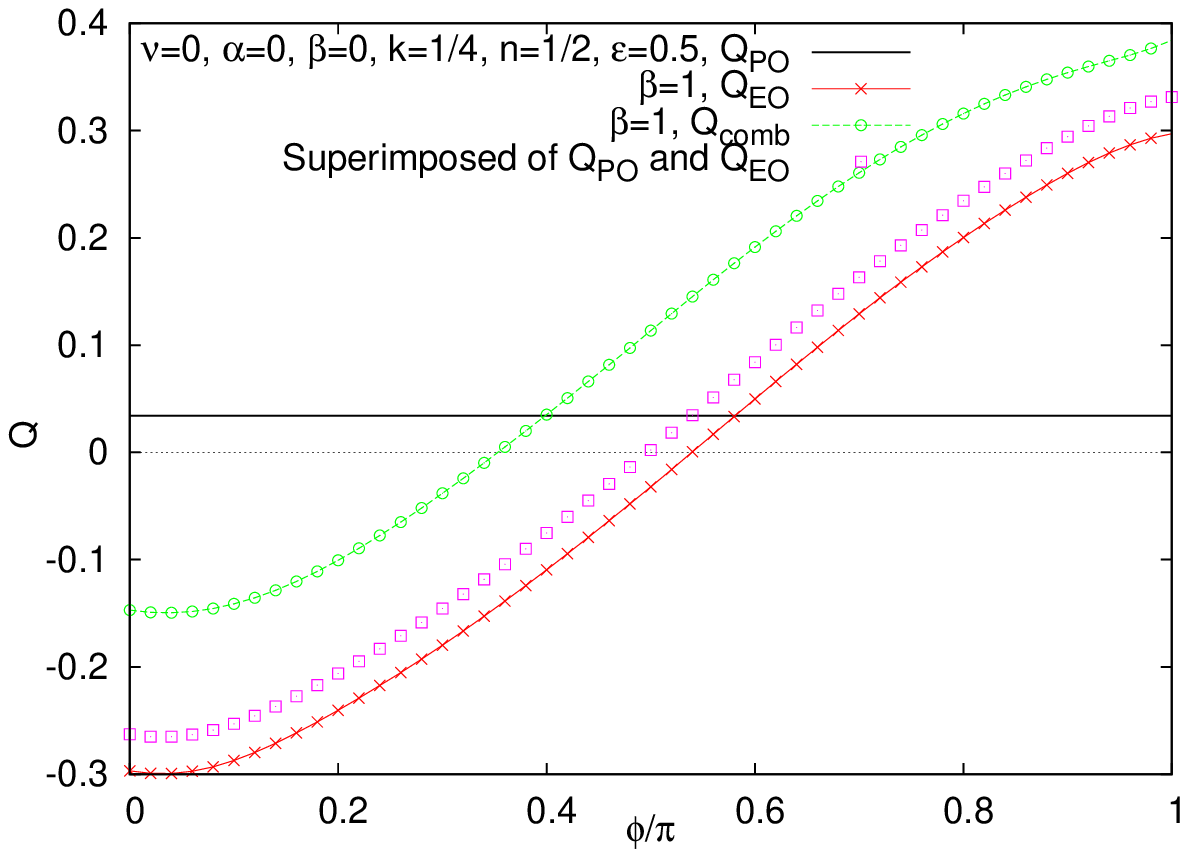}\includegraphics[width=3.5in,height=2.0in]{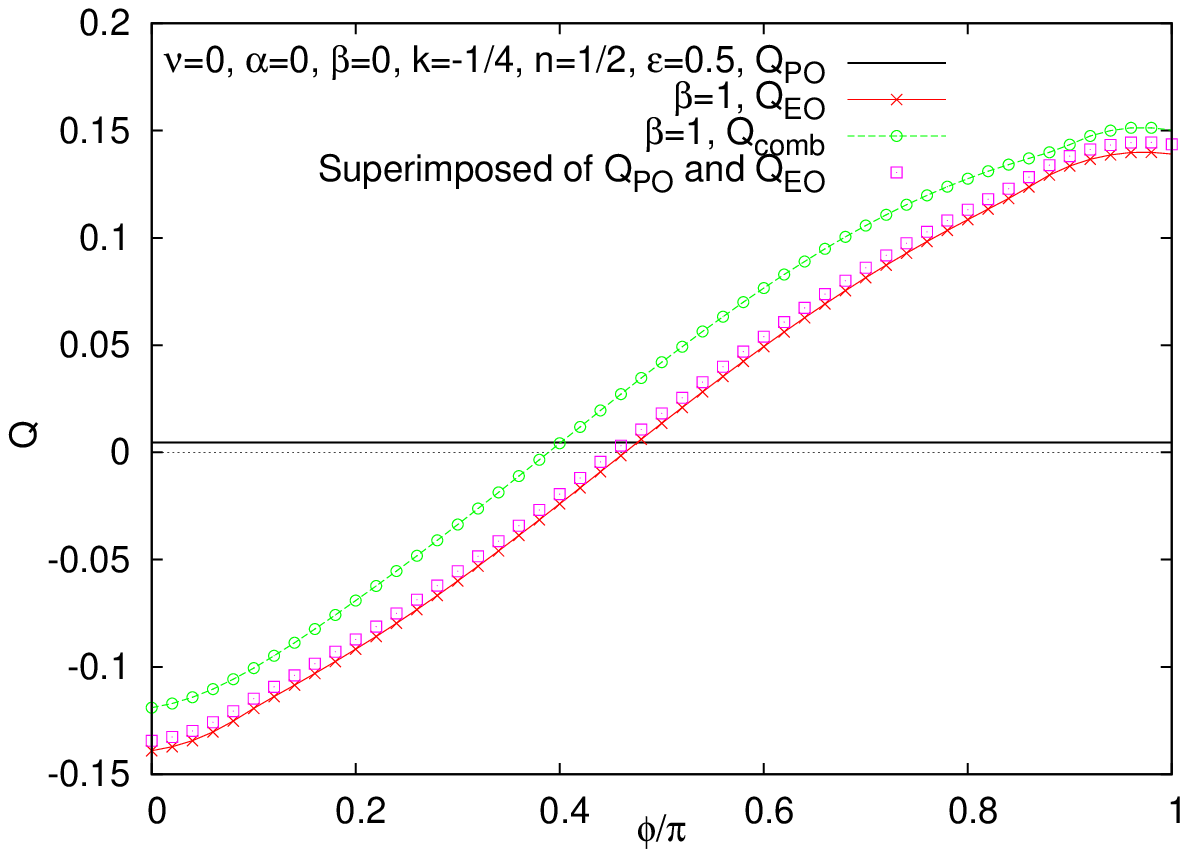}
\\$~~~~~~~~~~~~~~~~~(a)~~~~~~~~~~~~~~~~~~~~~~~~~~~~~~~~~~~~~~~~~~~~~~~~~~~~~~~~~~~~~~~~~~~(b)~~~~~~~~~$
\includegraphics[width=3.5in,height=2.0in]{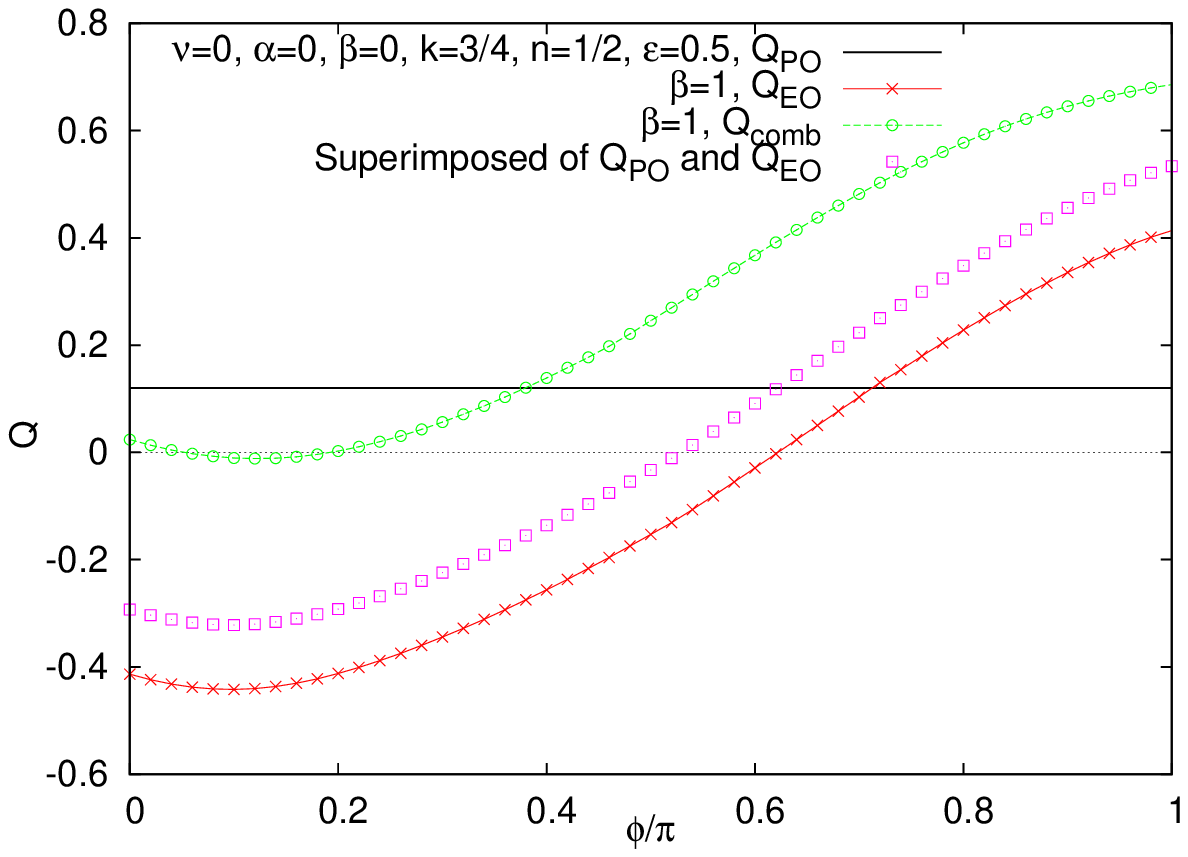}\includegraphics[width=3.5in,height=2.0in]{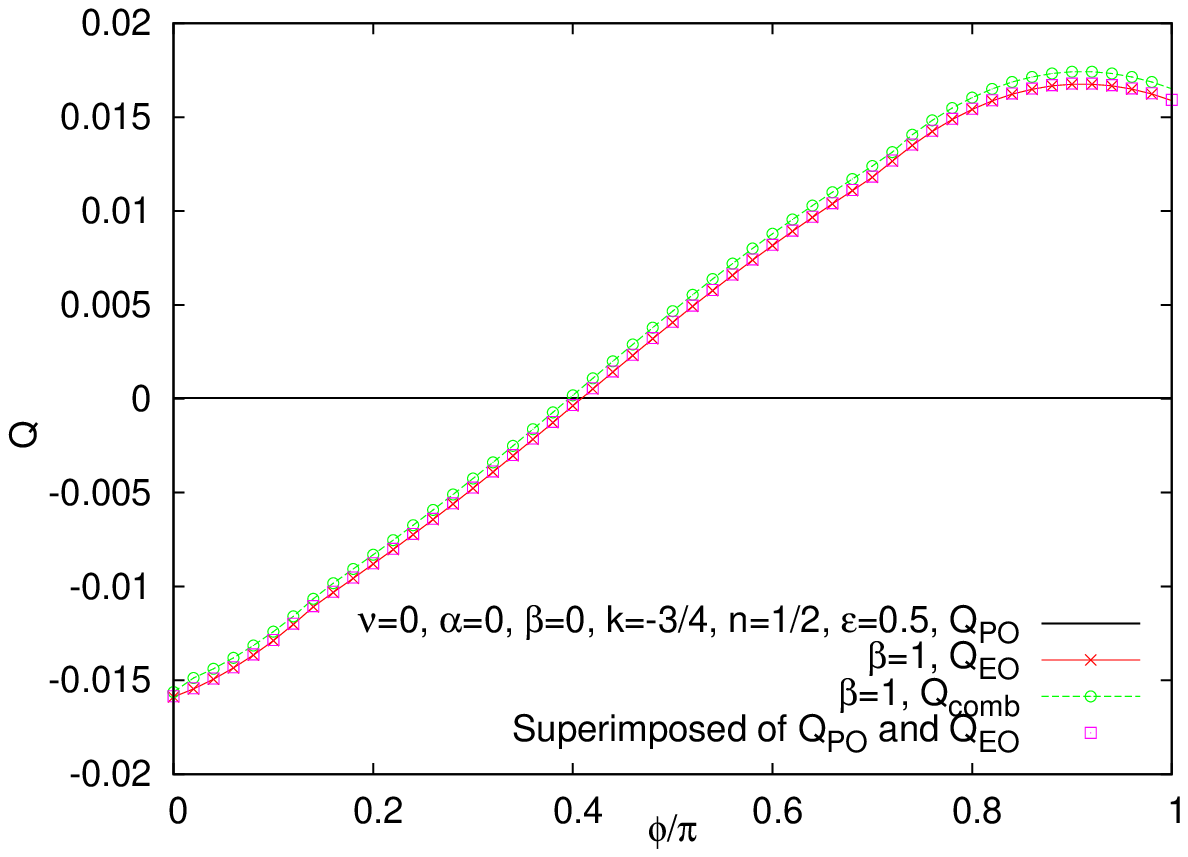}
\\$~~~~~~~~~~~~~~~~~(c)~~~~~~~~~~~~~~~~~~~~~~~~~~~~~~~~~~~~~~~~~~~~~~~~~~~~~~~~~~~~~~~~~~~(d)~~~~~~~~~$
\caption{Variation of flow rate $Q$ with the phase shift $\phi$ when
  $\Delta P=0,~-1;~\nu=0;~\epsilon=0.5;~\alpha=0;~\beta=0,~1$ for a
  shear-thinning fluid in tube (with mean radius is of diverging/tapered  nature).}
\label{manuscript_flow_rate_phase_diff_shear_thin_div_tap1.15-1.35}
\end{figure}

\begin{figure}
\includegraphics[width=3.5in,height=2.0in]{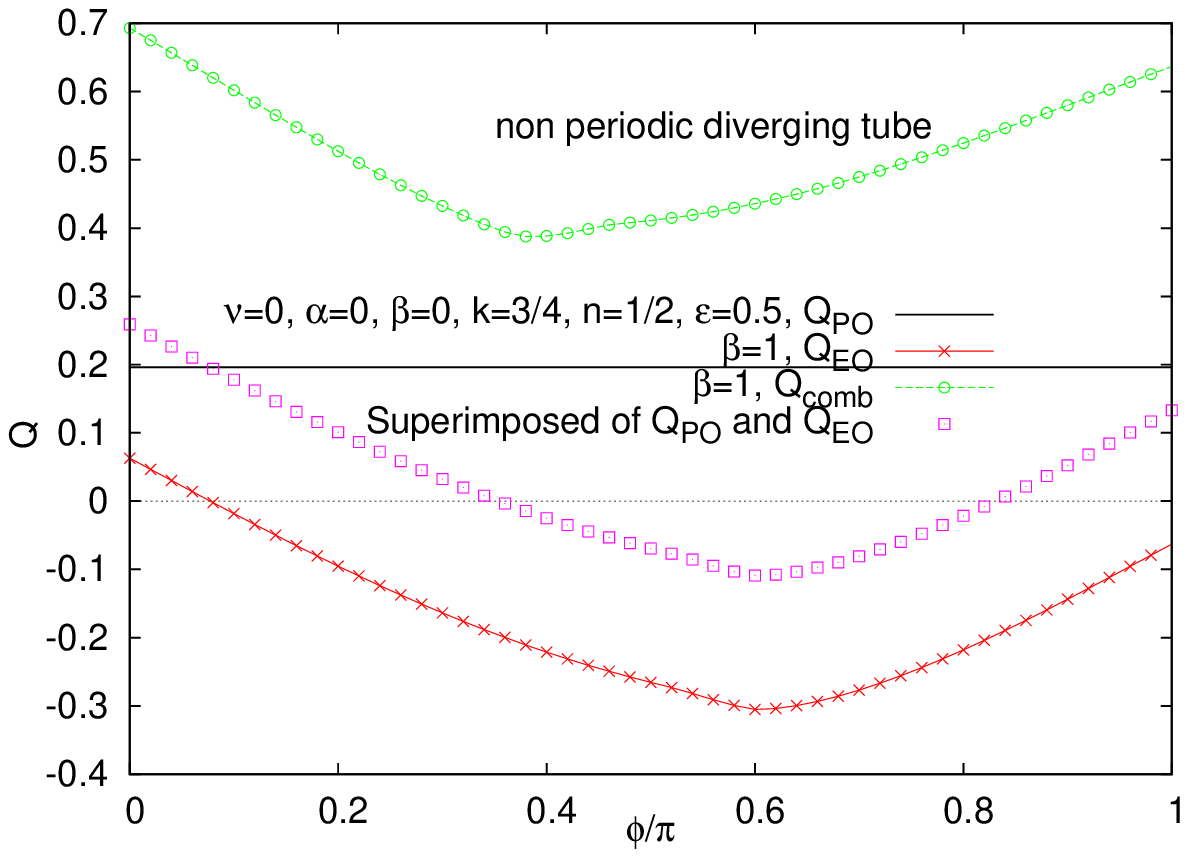}\includegraphics[width=3.5in,height=2.0in]{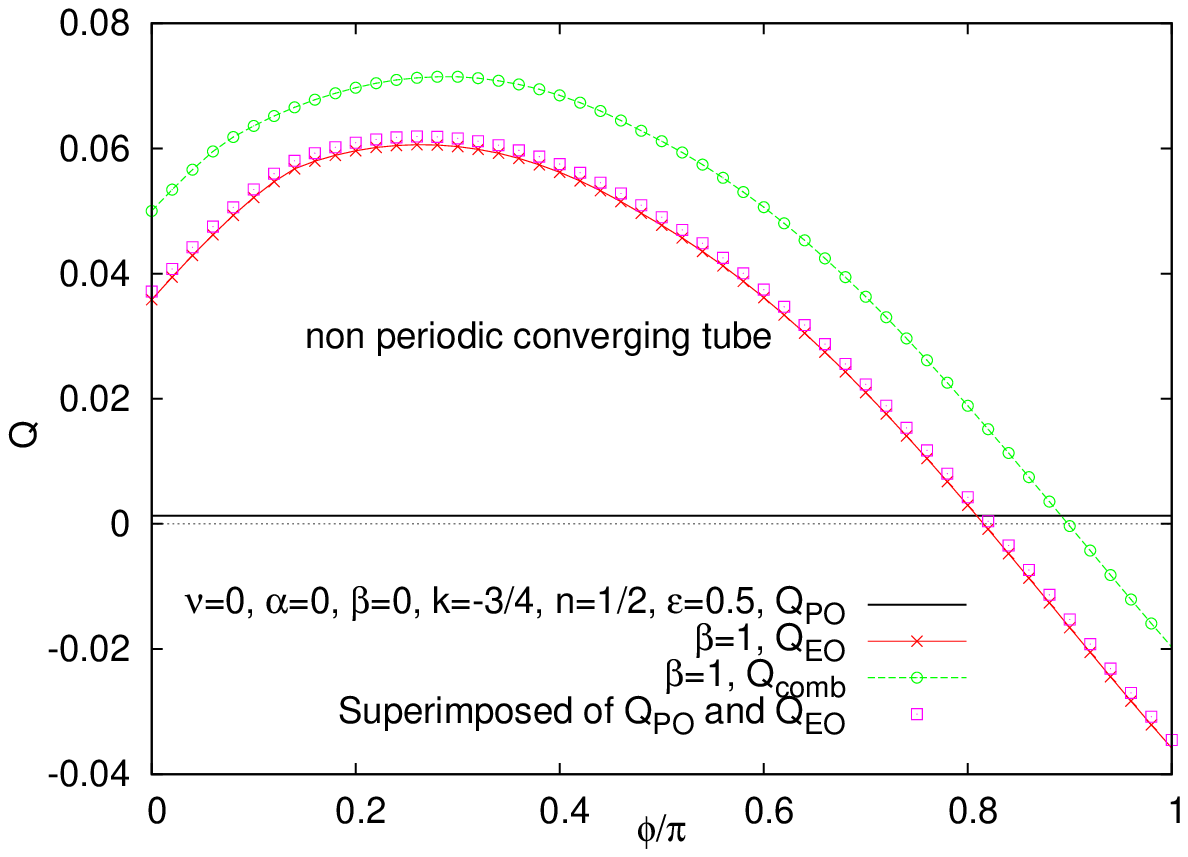}
\\$~~~~~~~~~~~~~~~~~(a)~~~~~~~~~~~~~~~~~~~~~~~~~~~~~~~~~~~~~~~~~~~~~~~~~~~~~~~~~~~~~~~~~~~(b)~~~~~~~~~$
\caption{Variation of flow rate $Q$ with the phase shift $\phi$ when
  $\Delta P=0,~-1;~\nu=0;~\epsilon=0.5;~\alpha=0;~\beta=0,~1$ for a
  shear-thinning fluid in a diverging/tapered tube.}
\label{manuscript_flow_rate_phase_diff_shear_thick_div_tap1.47-1.51}
\end{figure}

\begin{figure}
\centering
\includegraphics[width=3.5in,height=2.0in]{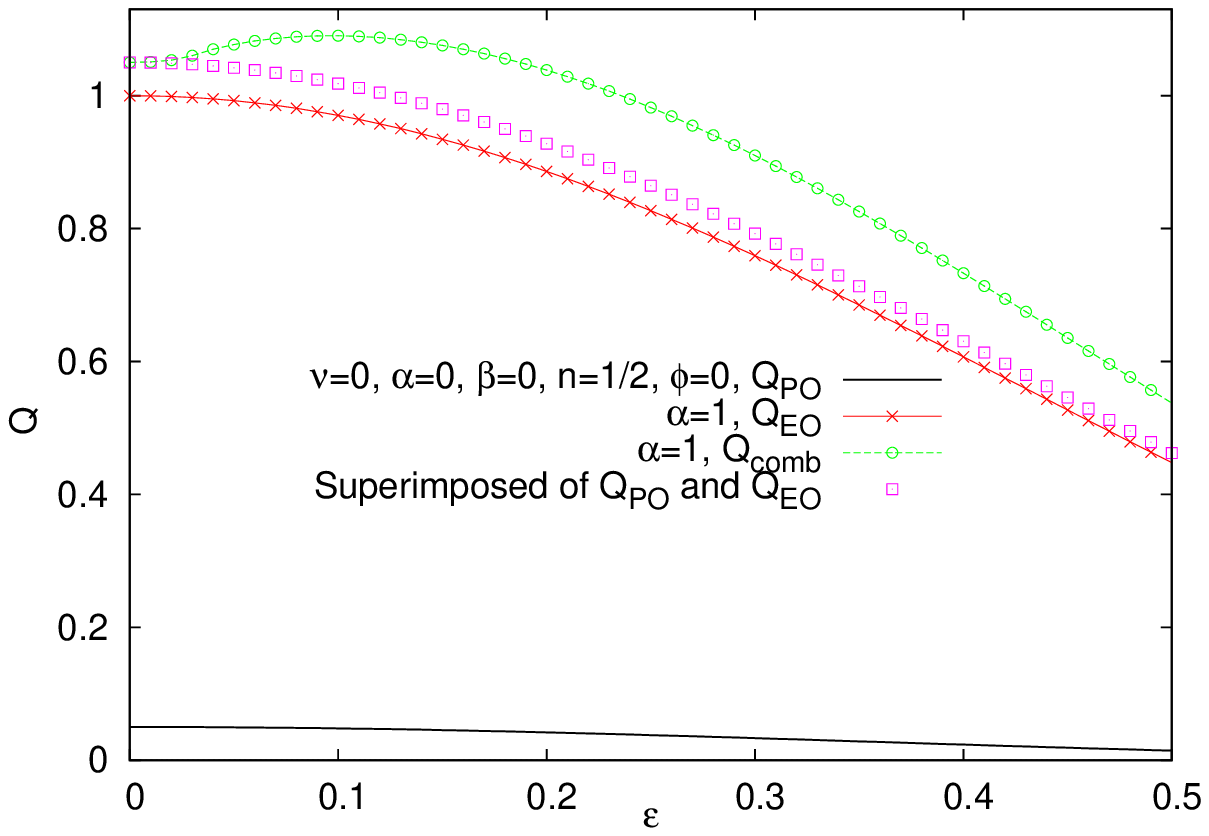}\includegraphics[width=3.5in,height=2.0in]{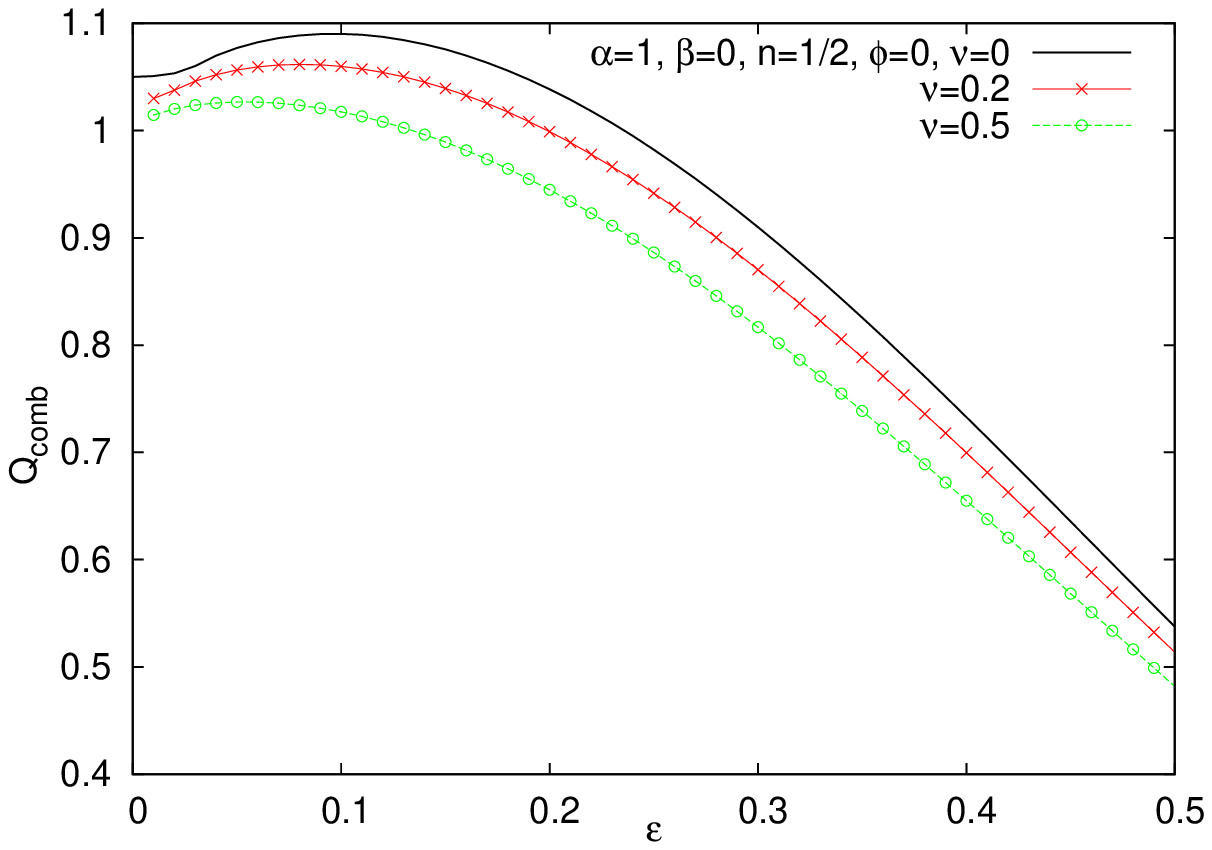}\\$~~~~~~~~~~~~~~~~~(a)~~~~~~~~~~~~~~~~~~~~~~~~~~~~~~~~~~~~~~~~~~~~~~~~~~~~~~~~~~~~~~~~~~~(b)~~~~~~~~~$
\includegraphics[width=3.5in,height=2.0in]{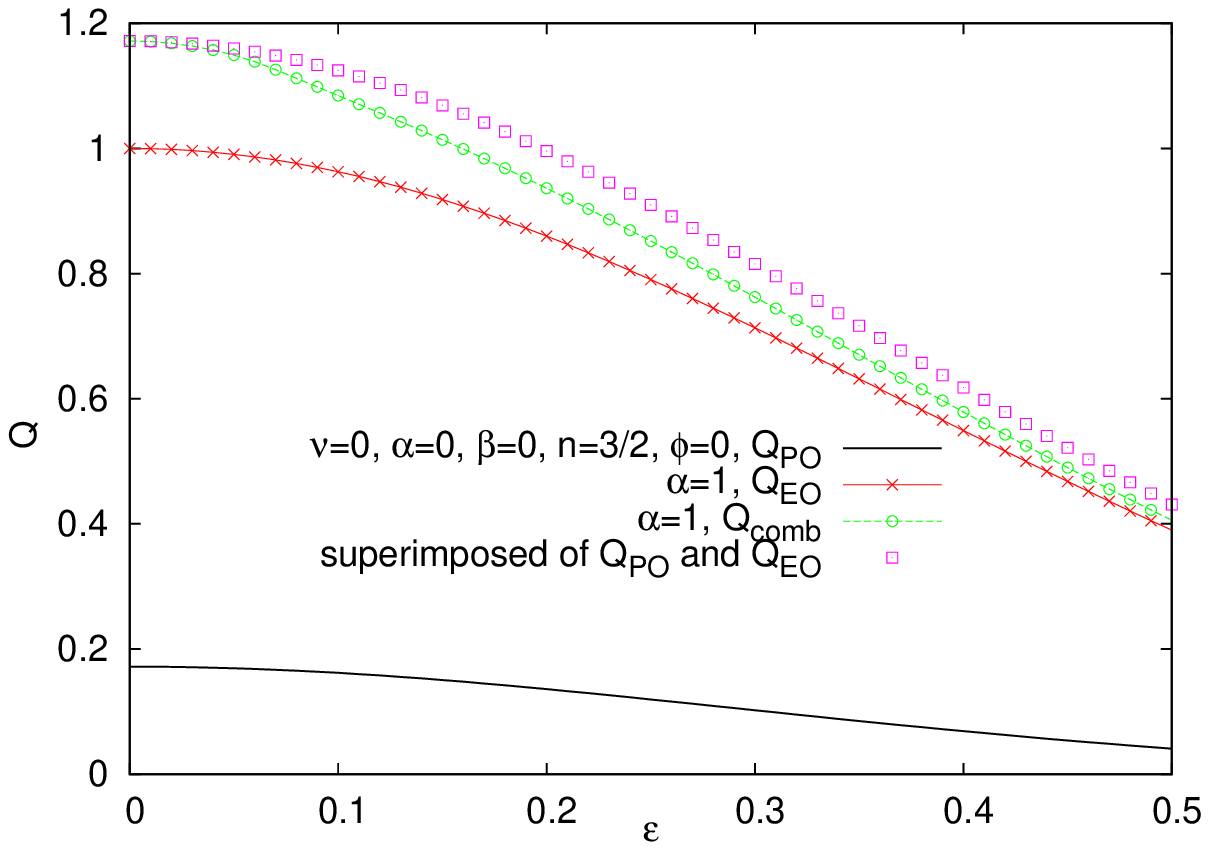}\includegraphics[width=3.5in,height=2.0in]{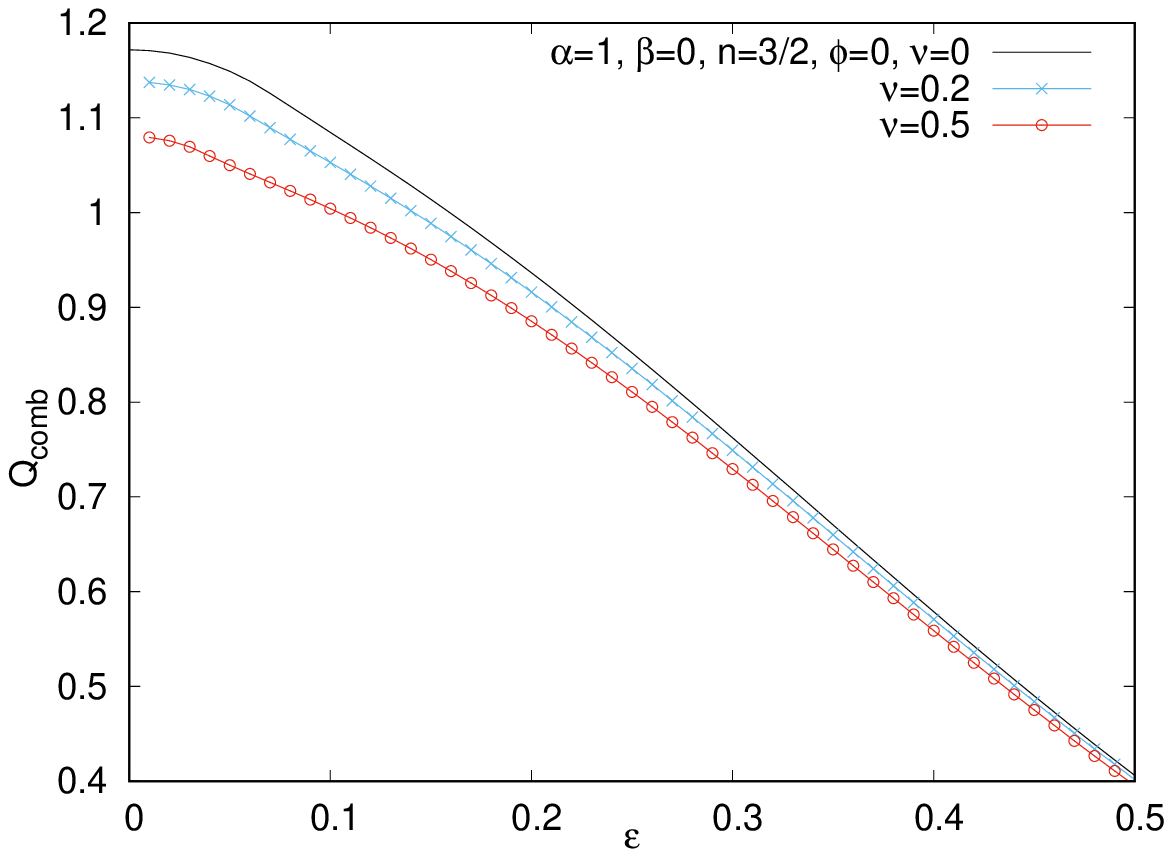}\\$~~~~~~~~~~~~~~~~~(c)~~~~~~~~~~~~~~~~~~~~~~~~~~~~~~~~~~~~~~~~~~~~~~~~~~~~~~~~~~~~~~~~~~~(d)~~~~~~~~~$
\caption{Variation of flow rate $Q$ with the wall undulation amplitude
  $\epsilon$ when $k=0;~\Delta
  P=0,-1;~\nu=0~\text{to }0.5;~\phi=0;~\alpha=0,~1;~\beta=0$ for a shear-thinning and a shear-thickening
  fluid.}
\label{manuscript_flow_rate_wall_amplitude_shear_thin4.1-4.4_shear_thick6.1-6.4}
\end{figure}

\begin{figure}
\centering
\includegraphics[width=3.5in,height=2.0in]{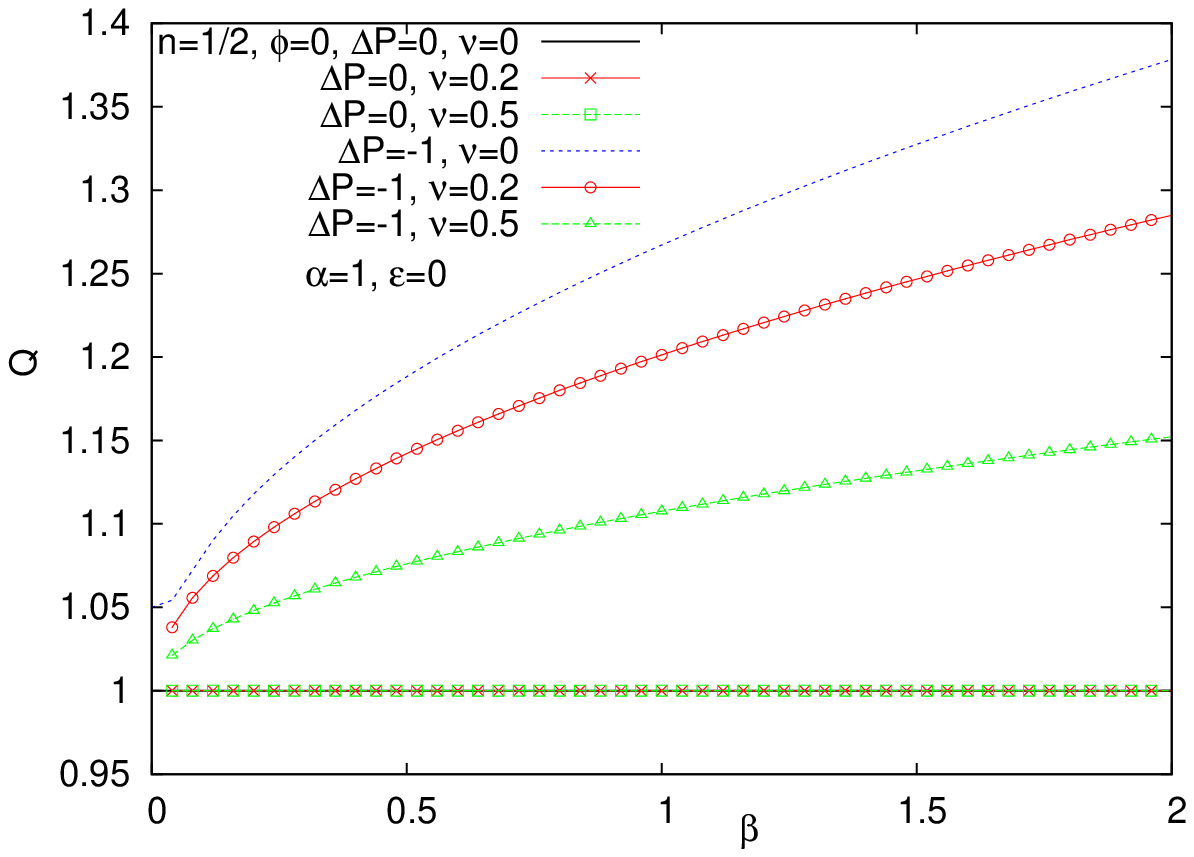}\includegraphics[width=3.5in,height=2.0in]{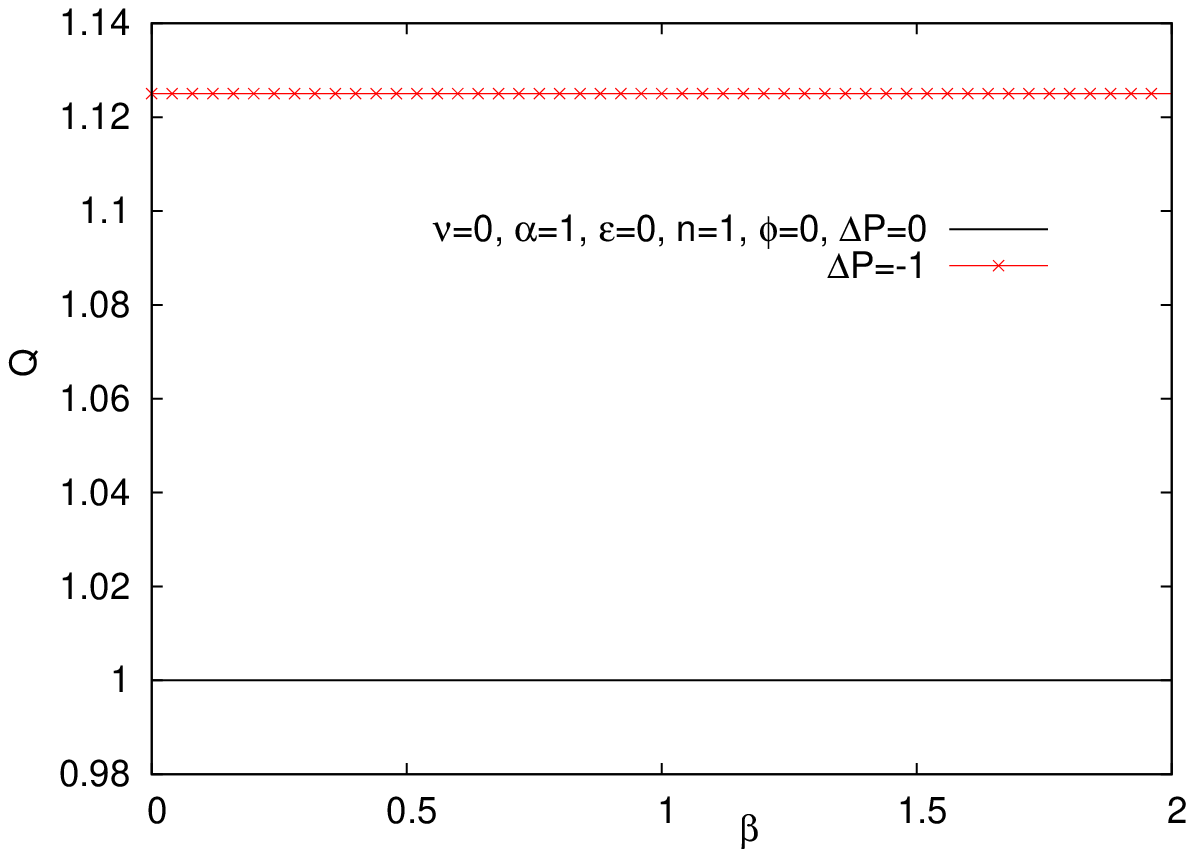}\\$~~~~~~~~~~~~~~~~~(a)~~~~~~~~~~~~~~~~~~~~~~~~~~~~~~~~~~~~~~~~~~~~~~~~~~~~~~~~~~~~~~~~~~~(b)~~~~~~~~~$
\includegraphics[width=3.5in,height=2.0in]{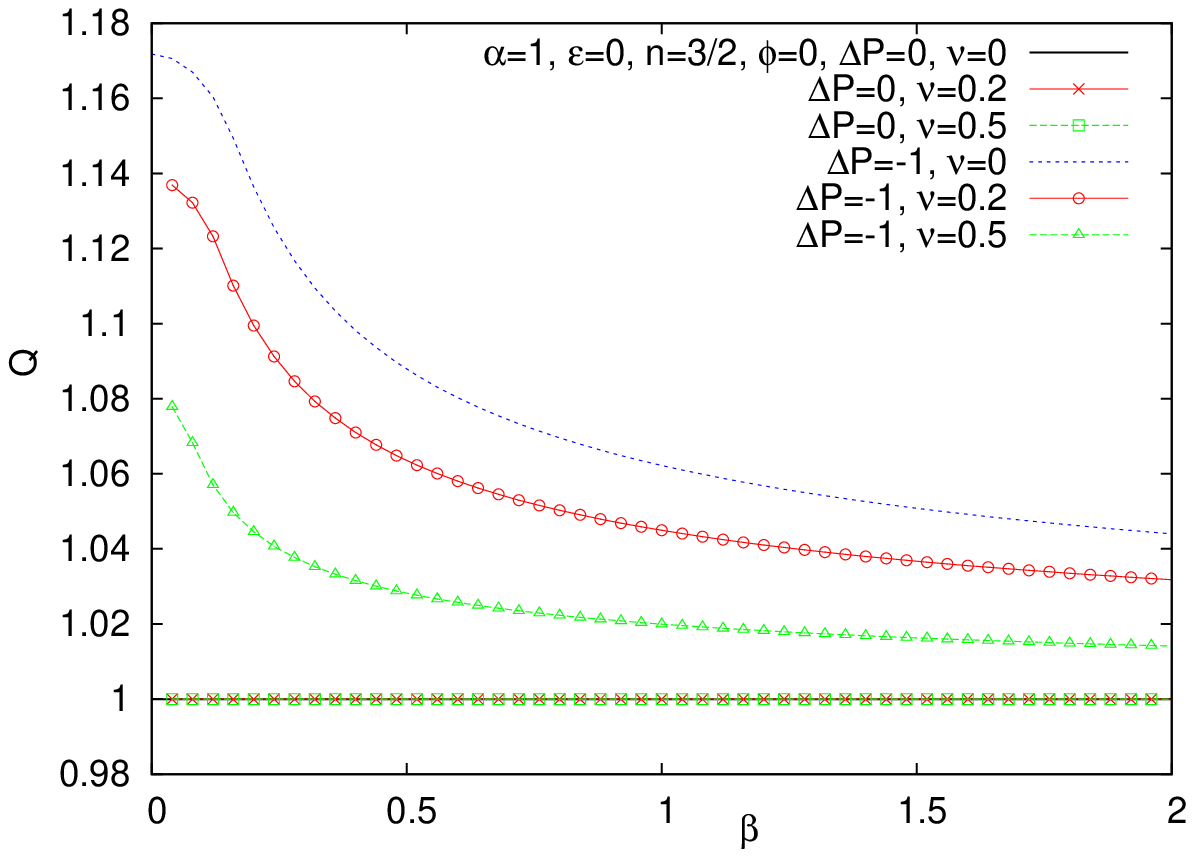}\\(c)
\caption{Variation of flow rate $Q$ with the HS slip modulation
  amplitude $\beta$ for $k=0;~\Delta
  P=0,-1;~\nu=0~\text{to }0.5;~\phi=0;~\alpha=1;~\epsilon=0$ (a) Newtonian fluid,
  (b) Shear-thinning fluid (c) Shear-thickening fluid.}
\label{manuscript_flow_rate_HS_slip7.2-9.2}
\end{figure}

\begin{figure}
\centering
\includegraphics[width=3.5in,height=2.0in]{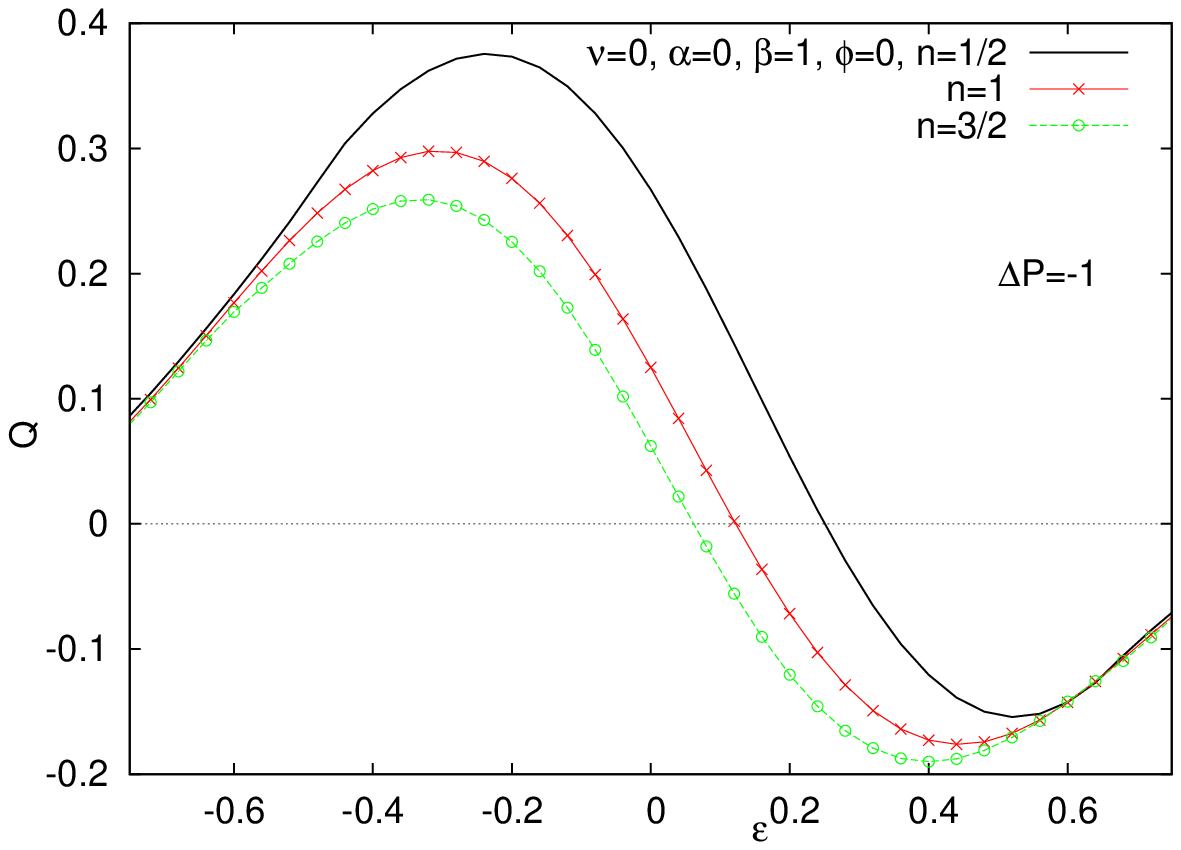}\includegraphics[width=3.5in,height=2.0in]{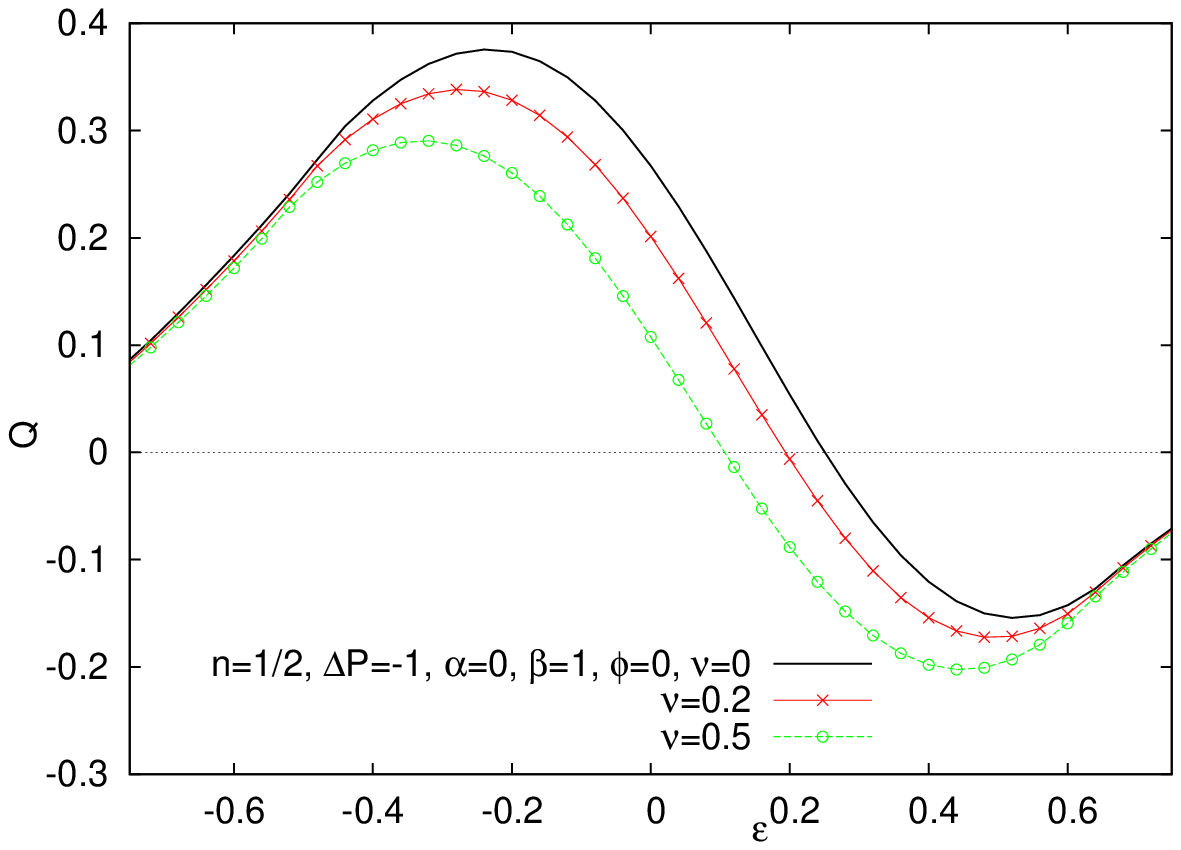}\\$~~~~~~~~~~~~~~~~~(a)~~~~~~~~~~~~~~~~~~~~~~~~~~~~~~~~~~~~~~~~~~~~~~~~~~~~~~~~~~~~~~~~~~~(b)~~~~~~~~~$
\includegraphics[width=3.5in,height=2.0in]{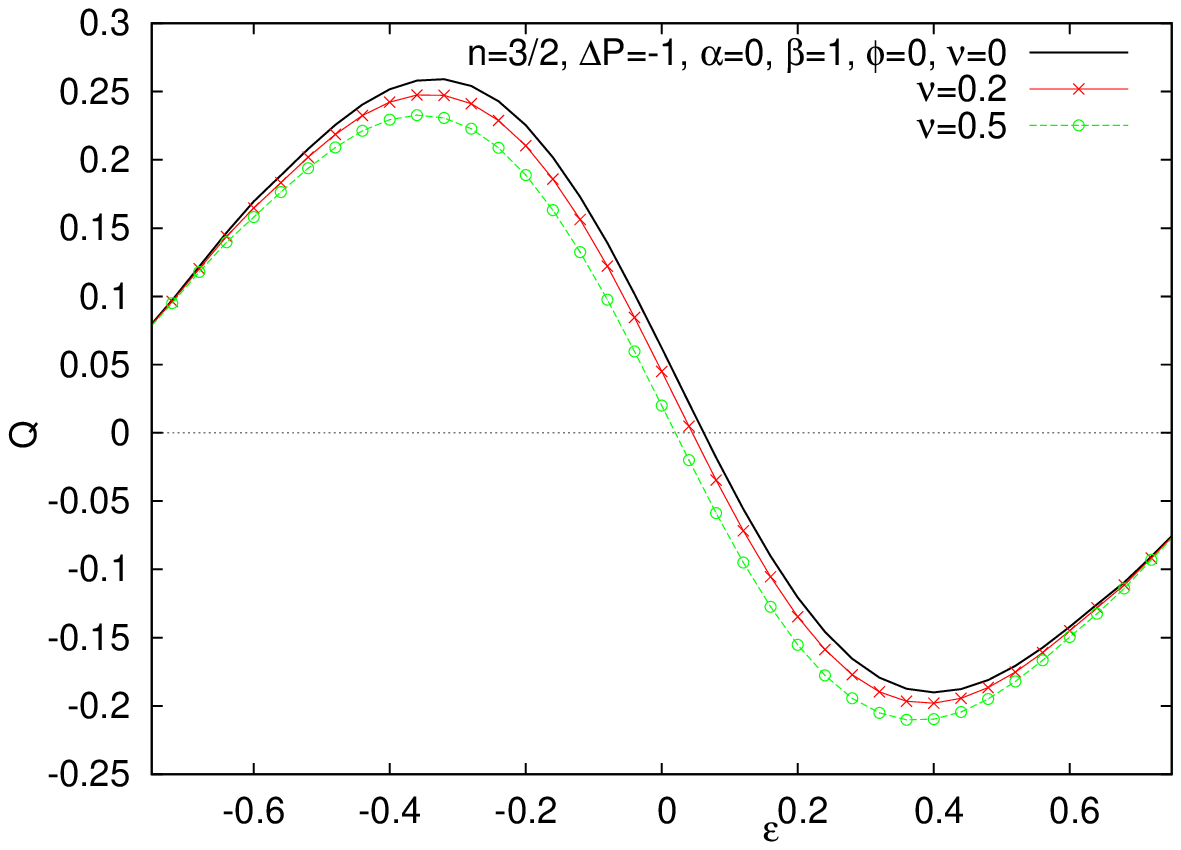}\\(c)
\caption{Variation of flow rate $Q$ with the wall undulation amplitude
  $\epsilon$ when $k=0;~\Delta P=-1,~\nu=0~\text{to }0.5,~\phi=0,~\alpha=0,~\beta=1$ for
  a Newtonian fluid, shear-thinning fluid and shear-thickening fluid.}
\label{manuscript_wall_slip_modulation_amplitude10.1-10.3}
\end{figure}

\begin{figure}
\centering
\includegraphics[width=3.5in,height=2.0in]{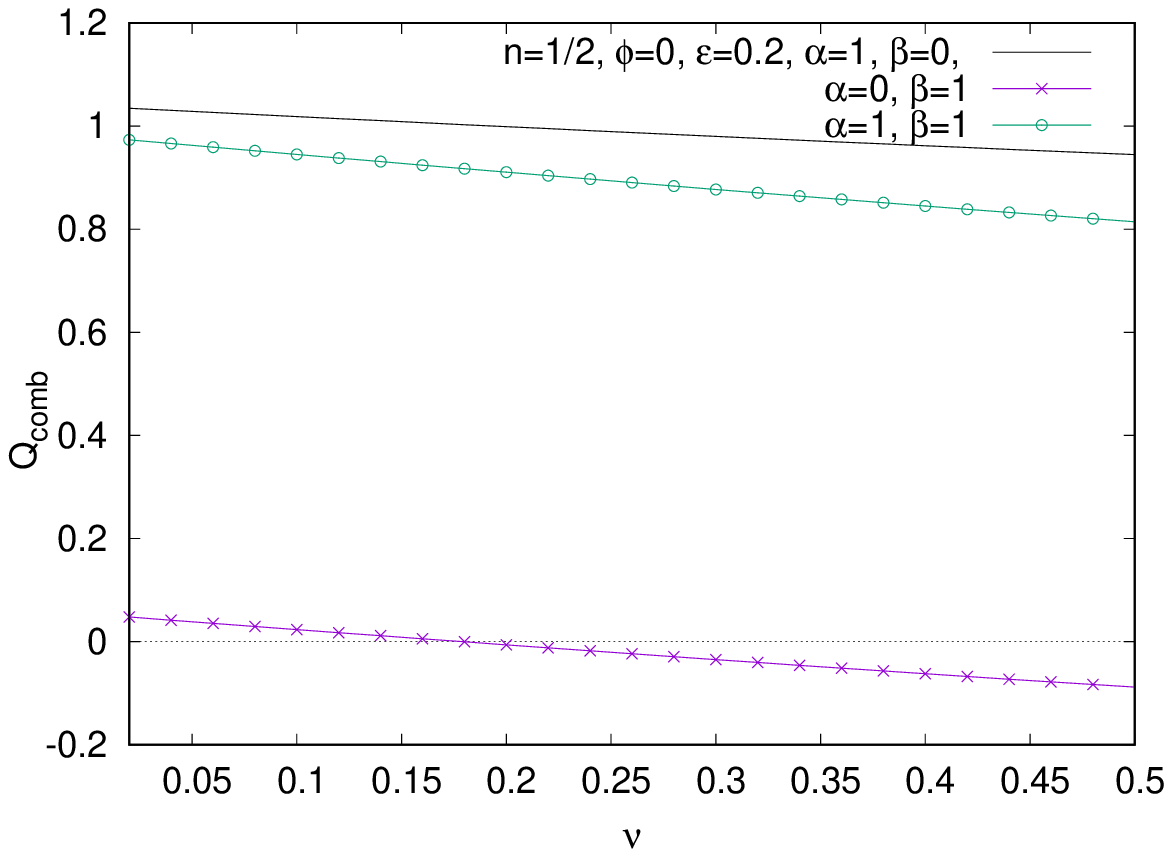}\includegraphics[width=3.5in,height=2.0in]{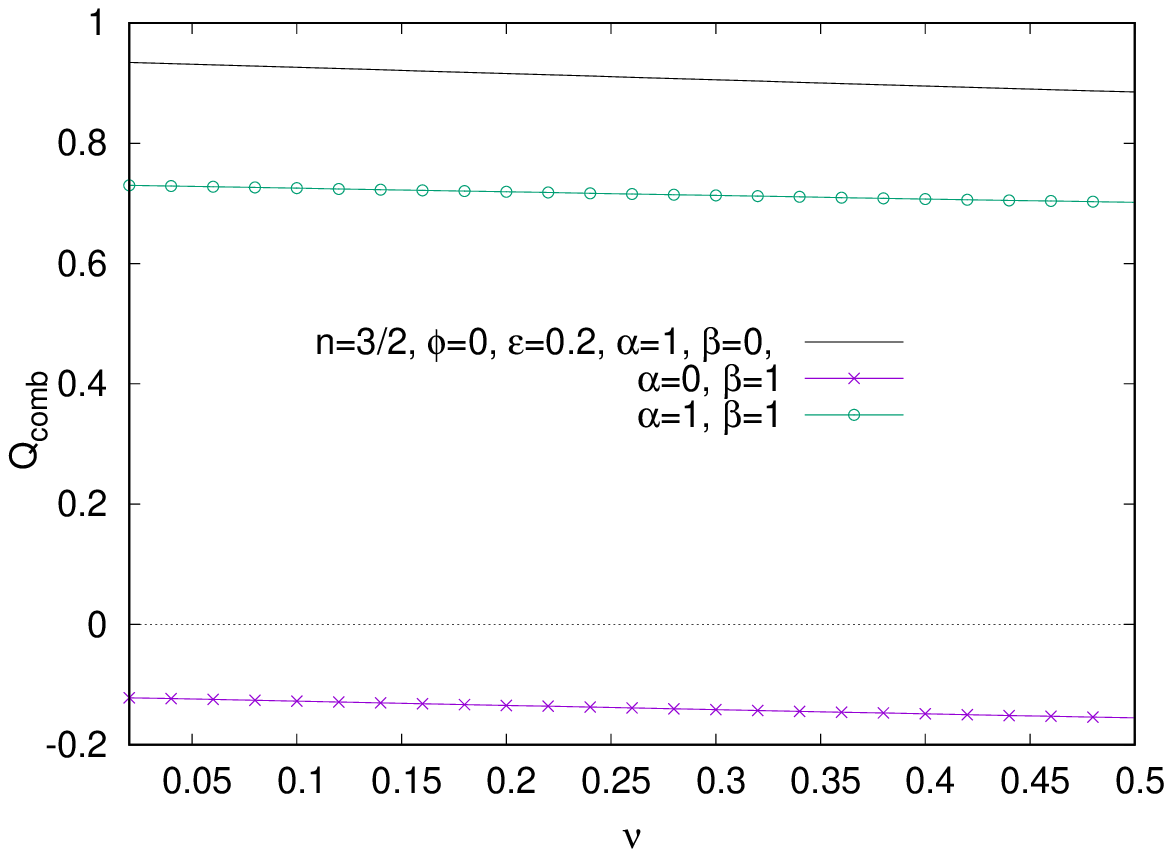}\\$~~~~~~~~~~~~~~~~~(a)~~~~~~~~~~~~~~~~~~~~~~~~~~~~~~~~~~~~~~~~~~~~~~~~~~~~~~~~~~~~~~~~~~~(b)~~~~~~~~~$
\caption{Variation of flow rate $Q_{comb}$ with parameter $\nu$ when
  $k=0,~\epsilon=0.2,~\phi=0,~\Delta P=-1$ (a) for a shear-thinning
  fluid ($n=1/2$), (b) for a shear-thickening fluid ($n=3/2$).}
\label{manuscript_flow_yield_stress_ratio51.1-51.4}
\end{figure}

\begin{figure}
\centering
\includegraphics[width=3.5in,height=2.0in]{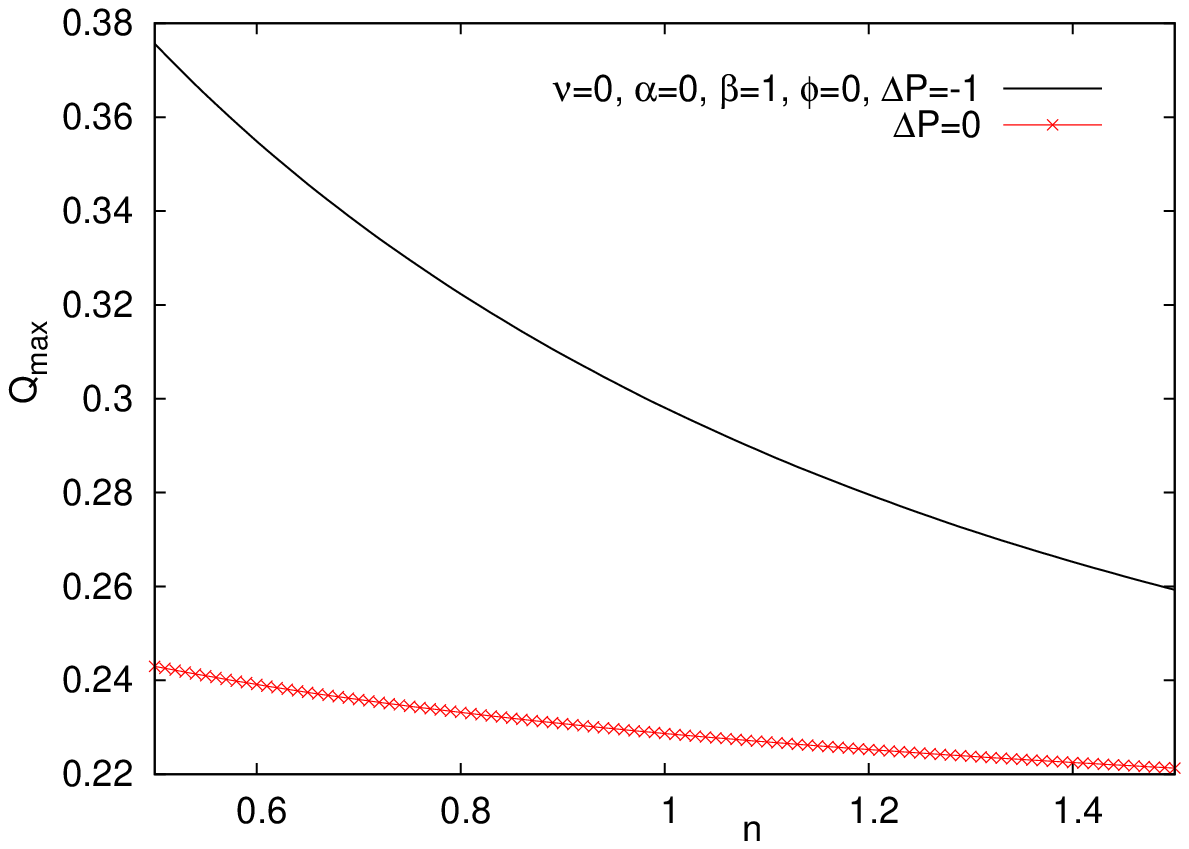}\includegraphics[width=3.5in,height=2.0in]{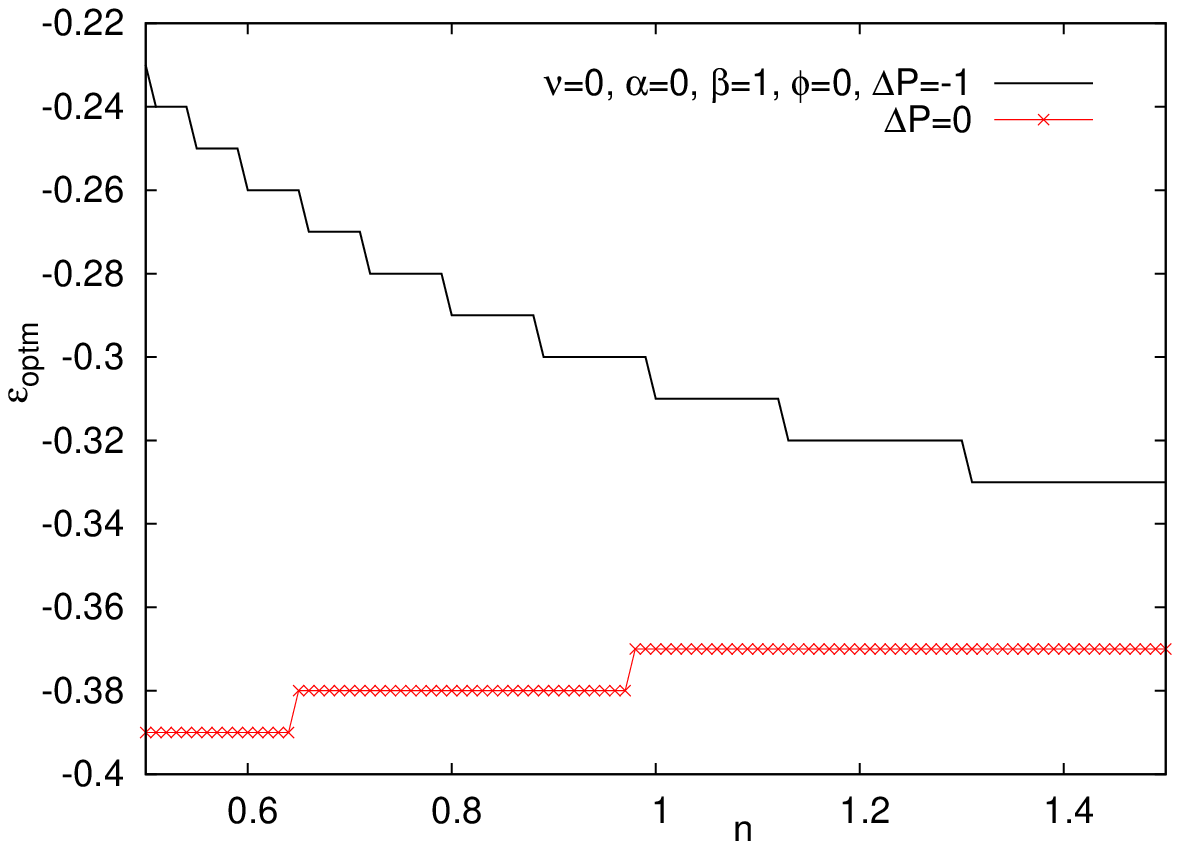}\\$~~~~~~~~~~~~~~~~~(a)~~~~~~~~~~~~~~~~~~~~~~~~~~~~~~~~~~~~~~~~~~~~~~~~~~~~~~~~~~~~~~~~~~~(b)~~~~~~~~~$
\caption{Variation of flow rate $Q_{max}$ and corresponding
  $\epsilon_{optm}$ with flow index number $n$ for
  $k=0,~\nu=0,~\phi=0,~\alpha=0,~\beta=1$ (a) for $Q_{max}$, (b) for
  $\epsilon_{optm}$.}
\label{manuscript_flow_max_wall_optm14.52-14.53}
\end{figure}

For different values of the phase shift $\phi$ (between the two
modulation waves), $n$ (flow index number), $\nu$, $\alpha$, $\beta$,
$\epsilon$,
Figs. \ref{manuscript_flow_rate_phase_diff_Newt2.1}-\ref{manuscript_flow_max_wall_optm14.52-14.53}
present the distribution of flow rate $Q$ for the cases of $Q_{PO}$,
$Q_{EO}$ and $Q_{comb}$. Figs.
\ref{manuscript_flow_rate_phase_diff_Newt2.1}-\ref{manuscript_flow_rate_phase_diff_shear_thick3.1-3.4}
show the the variation of flow rates with the phase $\phi$ for the
cases due to hydrodynamic forcing $\Delta P$ only, electric forcing
$U_s$ only and combined action of $\Delta P$ and $U_s$ for Newtonian
fluid ($n=1$), shear-thinning fluid ($n=1/2$) and shear-thickening
fluid ($n=3/2$). Moreover, the linear sum of hydrodynamic forcing
$Q_{PO}$ and electric forcing $Q_{EO}$ has been shown by the
rectangular points. It is worthwhile to note the significant influence
of the phase $\phi$ on the distributions of $Q_{PO}$, $Q_{EO}$ and
$Q_{comb}$ for the flows in non-uniform vessels. $Q_{EO}$ and
$Q_{comb}$ are enhanced gradually as the value of $\phi$ increases in
the interval $0<\phi<1$ except in the limiting case of $\phi=1$. It is
important to note that the flow is taking place in the backward
direction before $\phi$ reaches a certain value (near 0.4) in the
presence of an applied electric forcing. The sum of $Q_{PO}$ and
$Q_{EO}$ is different from $Q_{comb}$ at any phase shift for
shear-thinning fluid and shear-thickening fluid
(cf. Figs. \ref{manuscript_flow_rate_phase_diff_shear_thin1.1-1.4}-\ref{manuscript_flow_rate_phase_diff_shear_thick3.1-3.4}). However,
for Newtonian fluid
(cf. Fig. \ref{manuscript_flow_rate_phase_diff_Newt2.1}), $Q_{comb}$
is linear superposition of the hydrodynamic $Q_{PO}$ and
electrokinetic effects $Q_{EO}$. Thus for a non-Newtonian fluid, the
linearity fails to hold when $\epsilon$ and $\beta$ are non-zero. It
is interesting to note that $Q_{PO}+Q_{EO}<Q_{comb}$ for
shear-thinning fluid
(cf. Fig. \ref{manuscript_flow_rate_phase_diff_shear_thin1.1-1.4}),
but just opposite behaviour (i.e. $Q_{PO}+Q_{EO}>Q_{comb}$) is
observed for shear-thickening fluid
(cf. Fig. \ref{manuscript_flow_rate_phase_diff_shear_thick3.1-3.4}). If
either the radius of the tube or the wall charge distribution is
non-uniform, pressure can be generated internally. In order to
maintain the continuity of flow along the tube, this type of induced
pressure is required. For a power-law fluid, it is well known that the
flow rate varies non-linearly with the pressure gradient. However, for
EOF, the pressure gradient is generated not only due to the
hydrodynamic forcing, but also due to the electric forcing. As a
result, the relationship between the flow rate and the two forcings is
nonlinear for a non-Newtonian fluid flow in a tube of non-uniform
cross section. However, the wall undulation can interact with the wall
charge modulation differently for different values of the phase shift
$\phi$. It is further to be noted that the flow rate is the maximum
positive when $\phi=\pi$ and is the negative and minimum when
$\phi=0$. The phases $\phi=\pi$ and $\phi=0$ indicate the situations
where the maximum positive/negative slip velocity arises at the
narrowest section of the tube respectively. It also indicates that the
electric forcing at the narrowest part of the tube has the most
influence on the net flow. It further reveals the well known fact
consistently, i.e., the flow rate is limited at the smallest
cross-section of the channel/tube for the flow through a channel/tube
with transverse topographical patterns. Given wall shape and HS slip
modulation can interact with each other to give rise to the maximum
positive effect on the flow when $-\epsilon\beta\cos\phi$ is the
maximum positive. Hence $\phi=\pi$ if $\epsilon \beta >0$ and $\phi=0$
if $\epsilon\beta<0$. It is worthwhile to mention that similar results
are reported by Ng and Qi \cite{Ng} for their study of EOF of a
power-law fluid in a non-uniform micro-channel. However, the magnitude
of the above results (i.e. $Q_{PO}$, $Q_{EO}$ and $Q_{comb}$) for any
value of flow index number $n$ have been reduced significantly (around
half in magnitude) for EOF of a Herschel-Bulkley fluid in a tube of
non-uniform cross section.

Variation of $Q$ with respect to $\phi$ is affected by the value of
the parameter $\nu$ only if an applied pressure force is active
(cf. Figs. \ref{manuscript_flow_rate_phase_diff_shear_thin1.1-1.4},\ref{manuscript_flow_rate_phase_diff_shear_thick3.1-3.4}). The
effect is prominent when both the electric and pressure force are
acting
(cf. Figs. \ref{manuscript_flow_rate_phase_diff_shear_thin1.1-1.4}(d),\ref{manuscript_flow_rate_phase_diff_shear_thick3.1-3.4}(d)). The
backward flow has been enhanced and the forward flow has been reduced
with an increase in $\nu$ due to resistance created by the yield
stress. Moreover, $\nu$ has relatively more influence on
shear-thinning fluid than shear-thickening fluid.

Figs. \ref{manuscript_flow_rate_phase_diff_shear_thin_div_tap1.15-1.35}
exhibit the variation of flow rate $Q$ with the phase $\phi$ for the
cases due to $Q_{PO}$, $Q_{EO}$ and $Q_{comb}$ in a tube with mean
radius is of diverging/tapered nature. For the diverging case, the
flow rate is gradually and significantly increasing with the increase
of the inclination of the tube wall. Moreover, quantitative deviation
of $Q_{comb}$ from superimpose of $Q_{PO}$ and $Q_{EO}$ becomes more
prominent. For the converging case, the reverse trend of $Q$ has been
observed compare to previous case, i.e., $Q$ is gradually decreasing
at a similar percentage rate like the diverging case. However, when
outlet tube radius is 25$\%$ of the inlet radius, then hydrodynamic
force is negligible compare to the electrokinetic force and influence
of $\phi$ on $Q$ is very less impactful. Moreover, volumetric flow
($Q$) has been reduced significantly (near to zero) in this
case. These results may be expected physically as the resistance to
the flow decreases with the increases of free space in the tube for
the diverging case. Whereas, for the converging case the resistance to
the flow increases as the free space in the tube becomes
narrower. Thus, converging/diverging nature of the mean tube radius
plays vital role on the transport of a fluid.

In
Figs. \ref{manuscript_flow_rate_phase_diff_shear_thick_div_tap1.47-1.51},
a different nature has been observed for a tube where the tube radius
is of diverging/tapered nature (i.e. $h=1+kZ$ and $U_s=\alpha+\beta
\cos (2\pi Z+\phi)$) when it is compared to
Figs. \ref{manuscript_flow_rate_phase_diff_shear_thin_div_tap1.15-1.35}. In
this case, $\phi$ is considered as the phase constant which can be
interpreted using the initial amplitude at $Z=0$. Up to a certain limit
of $\phi$, $Q$ is decreasing with the increase of $\phi$ in the
diverging tube. If $\phi$ crosses the limit, then $Q$ is increasing
with the increase of $\phi$. It is to be noted that the minimum values
of $Q_{comb}$ and the superimpose of $Q_{PO}$ and $Q_{EO}$ have not
seen at the same limiting value of $\phi$ as hydrodynamic force and
electric force are not working linearly. We have found significant
deviation of $Q_{comb}$ from superimpose of $Q_{PO}$ and $Q_{EO}$
quantitatively for the case of a diverging tube. Since $Q_{EO}<0$ for
all values of $\phi$ except in the neighbourhood of 0, electric force
is working in the backward direction. In the case of a converging
tube, $Q$ is increasing with the increase of $\phi$ up to a certain
limit of $\phi$ and after then it is decreasing with the increase of
$\phi$. However, the magnitudes of $Q$ are found very less compare to
the case of converging tube. It is due to the resistance created by a
converging tube by reducing the area of tube.

The variation of flow rates with $\epsilon$ for the cases $Q_{PO}$,
$Q_{EO}$ and $Q_{comb}$ of Newtonian and rheological fluids in
uniformly charged walls ($\alpha=1,~\beta=0$), which are without any
phase (i.e. $\phi=0$), has been exhibited in
Figs. \ref{manuscript_flow_rate_wall_amplitude_shear_thin4.1-4.4_shear_thick6.1-6.4}. Also
for this case, the sum of $Q_{PO}$ and $Q_{EO}$ is different from
$Q_{comb}$ for Non-Newtonian fluid except in the limiting case of
$\epsilon=0$. These figure also support our earlier statement that for
any value of rheological parameter $n$, linearity works for a tube
with uniform cross section (i.e. $\epsilon=0,~\beta=0$). Here it is
also further noticed that $Q_{PO}+Q_{EO}<Q_{comb}$ for $n<1$ and
$Q_{PO}+Q_{EO}>Q_{comb}$ for $n>1$
(cf. Fig. \ref{manuscript_flow_rate_wall_amplitude_shear_thin4.1-4.4_shear_thick6.1-6.4}). It
has been observed that when $\beta=0$, $Q$ decreases monotonically
with increasing $\epsilon$
(cf. Figs. \ref{manuscript_flow_rate_wall_amplitude_shear_thin4.1-4.4_shear_thick6.1-6.4}). For
this case also, the magnitude of $Q_{PO}$, $Q_{EO}$ and $Q_{comb}$
have been reduced appreciably to the corresponding case in
micro-channel (cf. Ng and Qi \cite{Ng}) although these behaviours are
similar. Moreover, difference between $Q_{comb}$ and $Q_{PO}+Q_{EO}$
is clearly visible for a rheological fluid.

Like the previous case, variation of $Q$ with respect to $\epsilon$ is
significantly affected by the value of the parameter $\nu$ only if the
pressure force is acting
(cf. Figs. \ref{manuscript_flow_rate_wall_amplitude_shear_thin4.1-4.4_shear_thick6.1-6.4}). The
flow has been reduced clearly with an increase in $\nu$ due to the
resistance created by the yield stress.

Fig. \ref{manuscript_flow_rate_HS_slip7.2-9.2} displays the
distribution of flow rates of Newtonian and rheological fluids with
the HS slip modulation amplitude $\beta$ for $\phi=0,~\alpha=1$ in a
uniform tube. It reveals that the periodic variation of the HS slip
has no net effect on the flow of a Newtonian fluid if the wall
undulation is absent. This nature is irrespective of the pressure
forcing.

The response to the pressure forcing is very different in the case of
a shear-thinning and shear-thickening fluid. If the walls are flat and
there is no applied pressure forcing (i.e. $\Delta P=0$), the flow
rate is always unaffected by $\beta$ for any $n$ (cf. equation
(\ref{manuscript_Q_uniform_wall_fre_pump})). However, the contrary
result is found in the presence of an applied pressure forcing
(i.e. $\Delta P\neq$ not equal to zero). In this case, it is seen that
the flow rate for non-Newtonian fluid (i.e. $n$ not equal to 1)
significantly changed due to the change in $\beta$. For favorable
pressure gradient (i.e. $\Delta P<0$), the flow rate $Q$ increases
with $\beta$ for a shear-thinning fluid ($n<1$) and it decreases with
$\beta$ for a shear-thickening fluid
($n>1$). Fig. \ref{manuscript_flow_rate_HS_slip7.2-9.2} further
reveals that variation in the HS slip alone may interact non-linearly
with the hydrodynamic forcing for a rheological fluid even in the
absence of wall undulation (i.e. $\epsilon=0$).

Similar to the previous cases, variation of $Q$ with respect to
$\beta$ is observed to reduce appreciably with an increase in $\nu$
only if the pressure force is acting
(cf. Fig. \ref{manuscript_flow_rate_HS_slip7.2-9.2}). However, the
effect is more prominent here than the previous cases. Contrary to the
Fig. \ref{manuscript_flow_rate_HS_slip7.2-9.2}, $Q$ does not varying
monotonically with increasing $\epsilon$ when $\beta$ not equal to
zero
(cf. Fig. \ref{manuscript_wall_slip_modulation_amplitude10.1-10.3}). When
$\Delta P=-1~,\alpha=0,~\beta=1,~\phi=0$, Fig. 5 depicts how flow rate
may vary non-monotonically with $\epsilon$. This behaviour appears due
to the two competing effects.

It is observed that the area at the narrowest section decreases and
hence the flow is reduced in this region when $\epsilon$ increases in
magnitude. However, on the other hand, since the interaction between
the wall undulation and the slip modulation will become stronger, the
flow will be augmented. At the narrowest section, as the radius of the
tube tends to zero (i.e. at the limit $\epsilon=\pm 1$), the flow rate
$Q$ will also tend to zero
(Fig. \ref{manuscript_wall_slip_modulation_amplitude10.1-10.3}). When
$\epsilon$ becomes zero, the correlation term $\epsilon \beta=0$ and
so the resulting flow rate $Q$ is not the maximum. It is further
observed that the maximum flow rate $Q_{max}$ takes place at a
negative value of $\epsilon$. As mentioned earlier, interaction of
$\epsilon$ and $\beta$ will affect positively on $Q$ if
$\epsilon\beta\cos\phi$ negative only. As exhibited in
Fig. \ref{manuscript_wall_slip_modulation_amplitude10.1-10.3}, $Q$
becomes always positive for negative values of $\epsilon$ and $Q$ is
always negative for positive values of $\epsilon$. Similar to the
previous cases, it is seen that the variation of $Q$ with respect to
$\epsilon$ is reduced significantly if we increase $\nu$.

Fig. \ref{manuscript_flow_yield_stress_ratio51.1-51.4} displays the
distribution of flow rates $Q_{comb}$ of a shear-thinning fluid
($n=1/2$) and shear-thickening fluid ($n=3/2$) with parameter $\nu$
for $k=0,~\phi=0,~\epsilon=0.2$. The flow has been observed to reduce
with an increase in $\nu$ (i.e. consequently related to yield stress)
due to resistance created by the yield stress.  Here, we can also note
that the influence of yield stress and consequently $\nu$ has
relatively more effect on a shear-thinning fluid than a
shear-thickening fluid.

Fig. \ref{manuscript_flow_max_wall_optm14.52-14.53} presents how the
power-law index $n$ makes influence on the maximum flow rate $Q_{max}$
and the corresponding optimum undulation amplitude
$\epsilon_{optm}$. It is observed that $Q_{max}$ always decreases with
increasing $n$. $\epsilon_{optm}$ may rise or fall depending on
$\Delta P$. It is being reduced in magnitude with increasing $n$ if
the hydrodynamic force is active. However, $Q_{max}$ is seen to rise
with an increase in $n$ when hydrodynamic force is absent. Moreover,
the change of $\epsilon_{optm}$ for the case is not prominent to
$n$. These behaviours again point out the nonlinear interaction
between the hydrodynamic forcing and the electric forcing in driving
the flow of a rheological fluid through a tube of non-uniform
cross-section.

\section{Summary and Conclusion}
The present paper deals with a simplified study of the electrokinetic
flow of a Herschel-Bulkley fluid. The flow is considered in a tube
where the wall potential and tube radius is gradually varying along
the axis of the tube. We have treated the slip condition, representing
a thin Newtonian depletion layer enclosing a still thinner electric
double layer, as Helmholtz–Smoluchowski (HS) slip type. The effects of
the phase $\phi$, the wall undulation amplitude $\epsilon$, HS slip
modulation amplitude $\beta$, the power-law index $n$, yield stress
parameter $\nu$ (the ratio between yield stress and wall shear stress)
and the converging/diverging nature of the tube (i.e. $k$) on the
volumetric flow rate is investigated in different situations under the
purview of the lubrication theory. The plots present a clear view of
the qualitative and quantitative variation of various fluid mechanical
parameters. It is observed that pressure can be generated internally
under non-uniformity in either the wall charge or in the cross section
of the tube height, which will upset the linear relationship between
flow and applied electric force for a non-Newtonian fluid. The wall
undulation (i.e. $\epsilon$) and the surface charge modulation
(i.e. $\beta$) may interact with each other when pressure and electric
forces acting simultaneously. As a result, various nonlinear flow
behaviors appear. The study reveals that the flow is a linear
superimpose of the components due to the hydrodynamic and electric
forcings for a Non-Newtonian bulk fluid only when the tube is of
uniform cross section or when the flow is strictly
one-dimensional. However, how much pressure can be generated
internally and how the the hydrodynamic $\&$ electric forcings will
interact each other that depend strongly on the rheology of the fluid
(i.e. flow index number $n$), wall undulation and boundary slip
parameter. The effect of Newtonian depletion layer is to cause the
flow rate to be insignificant for the power-law index $n$ when the
pressure force is inactive. Because if $\Delta p=0$ (i.e. there is no
applied pressure forcing) and the tube radius is of uniform cross
section, then the pressure gradient is insignificant globally as well
as locally. Rheological fluid index $n$ related to hydrodynamic
  force, i.e., related to variation of pressure/pressure gradient.
Moreover, it is well known that the flow rate varies nonlinearly with
the pressure gradient for a power-law fluid. Thus $n$ has no
significant influence on volumetric flow due to constant like pressure
throughout the tube. As a result, a Non-Newtonian fluid acts Newtonian
not only near the wall of the tube, but also throughout the tube.

In spite of the Newtonian depletion layer, the power-law rheology is
appeared to be more influential on the flow in the presence of
pressure force. The periodic change in HS slip is found no effect on
the volumetric flow rate $Q$ if the pressure force is inactive and
$\epsilon=0$. However, in the presence of pressure force, the periodic
change of the wall potential has dissimilar effects on $Q$ depending
on $n=1$, $n<1$ and $n>1$. This behaviour occurs even when the wall
undulation is absent. In contrast, the effect of $\nu$ on the
volumetric flow rate $Q$ is found prominent only when the pressure
gradient is applied.

$Q$ is reducing gradually with increasing $\epsilon$ when
$\beta=0$. However, there exists an optimum amplitude at which the
flow rate is the maximum when $\beta>0$ and this maximum occurs where
$\epsilon_{optm}\beta\cos\phi<0$. The maximum flow rate and the
corresponding optimum undulation amplitude may vary appreciably with
the power-law index $n$. $n$ makes strong influence on the maximum
flow rate and the corresponding optimum undulation amplitude when the
pressure force is present. It is important to mention that the
converging/diverging nature of the tube radius plays a crucial role on
the EOF of a rheological fluid.

There are many microfluidic processes involving various non-Newtonian
Power-law fluids. In order to dealing these fluids, the Smoluchowski
velocity may be applied in order to predict the flow rates of the
Power-law fluids in electrokinetically driven microfluidic
devices. Since microstructures in electrokinetics-based microfluidic
devices are complex, exact solutions are not available there. If we
apply the slip velocity approach with the Smoluchowski velocity as the
slip wall velocity, the numerical simulations may be
simplified. However, there are the limitations for the applying of
this assumption and also about the limitations on the validation of
results as discussed in the introduction and formulation of the study.

\vspace{0.5cm}

{\bf Acknowledgment:} {\it We thank the Reviewers and Associate
    Editor, Professor I. A. Frigaard for their useful comments, which
    helped to improve the paper. One of the authors, S. Maiti, is
    thankful to the University Grants Commission (UGC), New Delhi for
    awarding the Dr. D. S. Kothari Post Doctoral Fellowship during
    this investigation.}

\end{document}